\pdfoutput=1
\PassOptionsToPackage{hyphens}{url} %

\documentclass[5p,preprint]{elsarticle}

\journal{\href{https://www.journals.elsevier.com/journal-of-systems-and-software}{Journal of Systems and Software (JSS)}. Published version \doi{10.1016/j.jss.2020.110708}.}

\usepackage[english]{babel}
\usepackage[utf8x]{inputenc}
\usepackage[T1]{fontenc}
\usepackage{url} %
\usepackage{hyperref} %
\usepackage[usenames,dvipsnames,svgnames,table]{xcolor} %
\usepackage{graphicx} %
\usepackage{amsmath} %
\usepackage[strict]{changepage} %
\usepackage{listings} %
\usepackage{xspace} %
\usepackage{enumitem} %
\usepackage{booktabs} %
\usepackage[normalem]{ulem}
\useunder{\uline}{\ul}{}

\usepackage{microtype} %
\usepackage{siunitx} %
\usepackage{csquotes} %
\usepackage{doi} %

\usepackage{float} %
\usepackage{listings}

\newfloat{lstfloat}{htbp}{lop}
\floatname{lstfloat}{Listing}
\usepackage[colorinlistoftodos]{todonotes} %
\usepackage[switch]{lineno} %
\runningpagewiselinenumbers

\definecolor{plotfirstblue}{HTML}{6794a7}
\definecolor{plotbackground}{HTML}{d5e4eb}
\definecolor{plotdarkblue}{HTML}{014d64}
\definecolor{plotlightblue}{HTML}{7ad2f6} %
\definecolor{plotlightbluelighter}{HTML}{d9f2fc} %
\definecolor{plotmediumblue}{HTML}{01a2d9}

\usepackage{myrq}

\newcommand{\code}[1]{\texttt{\small #1}}
\newcommand{\footurl}[1]{\footnote{\url{#1}}}
\newcommand{\ie}{i.e.,\xspace}

\newcommand{\appendixurl}{https://github.com/joe4dev/faas-performance-mlr}

\newcounter{acounter}
\newcommand{\alabel}[1]{\refstepcounter{acounter}\label{#1}}

\newcounter{gcounter}
\newcommand{\glabel}[1]{\refstepcounter{gcounter}\label{#1}}

\begin{document}
\begin{frontmatter}

\title{Function-as-a-Service Performance Evaluation:\\
    A Multivocal Literature Review}

\author[chalmers]{Joel Scheuner}\corref{cor1}
\ead{scheuner@chalmers.se}

\author[chalmers]{Philipp Leitner}
\ead{philipp.leitner@chalmers.se}

\cortext[cor1]{Corresponding author}

\address[chalmers]{Software Engineering Division, Chalmers | University of Gothenburg, Sweden}

\begin{abstract}
Function-as-a-Service (FaaS) is one form of the serverless cloud computing paradigm and is defined through FaaS platforms (e.g., AWS Lambda) executing event-triggered code snippets (i.e., functions).
Many studies that empirically evaluate the performance of such FaaS platforms have started to appear but we are currently lacking a comprehensive understanding of the overall domain.
To address this gap, we conducted a multivocal literature review (MLR) covering 112 studies from academic (51) and grey (61) literature.
We find that existing work mainly studies the AWS Lambda platform and focuses on micro-benchmarks using simple functions to measure CPU speed and FaaS platform overhead (i.e., container cold starts).
Further, we discover a mismatch between academic and industrial sources on tested platform configurations, find that function triggers remain insufficiently studied, and identify HTTP API gateways and cloud storages as the most used external service integrations.
Following existing guidelines on experimentation in cloud systems, we discover many flaws threatening the reproducibility of experiments presented in the surveyed studies.
We conclude with a discussion of gaps in literature and highlight methodological suggestions that may serve to improve future FaaS performance evaluation studies.

\end{abstract}


\end{frontmatter}

\section{Introduction}\label{sec:intro}

Cloud computing continues to evolve, moving from low-level services such as Amazon Web Services (AWS) EC2, towards integrated ecosystems of specialized high-level services.
Early Infrastructure-as-a-Service (IaaS) cloud services are generalist solutions, which only provide a low-level abstraction of computing resources, typically in the form of self-administered virtual machines.
In contrast, the emerging serverless\footurl{https://martinfowler.com/articles/serverless.html} paradigm aims to liberate users entirely from operational concerns, such as managing or scaling server infrastructure, by offering a fully-managed high-level service with fine-grained billing~\citep{eyk:17}.
As a type of specialist service, serverless offerings range from simple object storage (e.g., Amazon S3) to deep learning-powered conversational agents (e.g., Amazon Lex, the technology behind Alexa).

To connect the different services (e.g, feed images from S3 into a transcoding service), a serverless-but-generalist service is required as a `glue' to bridge the gaps (in triggers, data formats, etc.) between services.
This is the primary niche that Function-as-a-Service (FaaS) platforms, such as AWS Lambda\footnote{\url{https://aws.amazon.com/lambda/}}, have emerged to fill.

In FaaS, developers provide small snippets of source code (often JavaScript or Python) in the form of programming language functions adhering to a well-defined interface.
These functions can be connected to trigger events, such as incoming HTTP requests, or data being added to a storage service.
The cloud provider executes the function (with the triggering event as input) on-demand and automatically scales underlying virtualized resources to serve elastic workloads of varying concurrency.
FaaS is used for a wide variety of tasks~\citep{leitner:19}, including as a `glue' holding together a larger serverless application, as a backend technology to implement REST services, and for a variety of data analytics (e.g., PyWren~\citep{jonas:17}) and machine learning tasks (e.g., serving deep learning models~\citep{ishakian:18}).
This makes their performance crucial to the efficient functioning of a wide range of cloud applications.

Previous research has indicated performance-related challenges common to many FaaS platforms. %
Among others, cold start times (the time required to launch a new container to execute a function) can lead to execution delays of multiple seconds~\citep{manner:18}, hardware heterogeneity makes predicting the execution time of a function difficult~\citep{figiela:18}, and complex triggering mechanisms can lead to significant delays in function executions on some platforms~\citep{pelle:19}.
So far, reports about performance-related challenges in FaaS are disparate and originate from different studies, executed with different setups and different experimental assumptions.
The FaaS community is lacking a consolidated view on the state of research on FaaS performance.

This paper addresses this gap.
We conduct a multivocal literature review (MLR)~\citep{garousi:19} to consolidate academic and industrial (i.e., grey literature) sources that were published between 2016 and 2019 and report performance measurements of FaaS offerings of different platforms.
The area of our study is the performance evaluation (also referred to as performance benchmarking) of FaaS offerings, both of commercial public services and open source systems intended to be installed in private data centers.
Our research goal is two-fold.
Firstly, we characterize the landscape of existing isolated FaaS performance studies.
Secondly, we identify gaps in current research (and, consequently, in our understanding of FaaS performance).
We also provide methodological recommendations aimed at future FaaS performance evaluation studies.

The remainder of this paper is structured as follows.
Section~\ref{sec:bg} introduces FaaS performance benchmarking.
Section~\ref{sec:rq} defines and motivates our research questions.
Section~\ref{sec:study} describes our MLR study design before we present and discuss the results in Section~\ref{sec:results}.
The main findings then lead to the implications of our study in Section~\ref{sec:gap}, where we also identify gaps in current literature.
Section~\ref{sec:relatedwork} relates our work and results to other research in the field.
Finally, Section~\ref{sec:conclusion} summarizes and concludes this paper.

\section{Background}\label{sec:bg}

This section introduces FaaS performance benchmarking based on the two benchmark types covered in this paper.
Micro-level benchmarks target a narrow performance aspect (e.g., floating-point CPU performance) with artificial workloads, whereas application-level benchmarks aim to cover the overall performance (i.e., typically end-to-end response time) of real-world application scenarios.
We clarify this distinction of benchmark types based on example workloads from our analyzed studies.

\subsection{Micro-Benchmarks}
Listing~\ref{lst:sample} shows a simple CPU-intensive AWS Lambda function written in the Python programming language.
This example function serves as a CPU micro-benchmark in one of our surveyed studies~[\ref{a21}].
It implements a provider-specific handler function to obtain the parameter \code{n} from its triggering invocation event (see line 13).
The floating point operations helper function (see line 4) exemplifies how common FaaS micro-benchmarks measure latency for a series of CPU-intensive operations.

\begin{lstfloat}
\begin{lstlisting}[label=lst:sample, caption=Lambda function with CPU micro-benchmark~[\ref{a21}]]
import math
from time import time

def float_operations(n):
    start = time()
    for i in range(0, n):
        sin_i = math.sin(i)
        cos_i = math.cos(i)
        sqrt_i = math.sqrt(i)
    latency = time() - start
    return latency

def lambda_handler(event, context):
    n = int(event['n'])
    result = float_operations(n)
    print(result)
    return result
\end{lstlisting}
\end{lstfloat}

\subsection{Application-Benchmarks}
\begin{figure}[htb]
    \centering
    \includegraphics[width=0.47\textwidth]{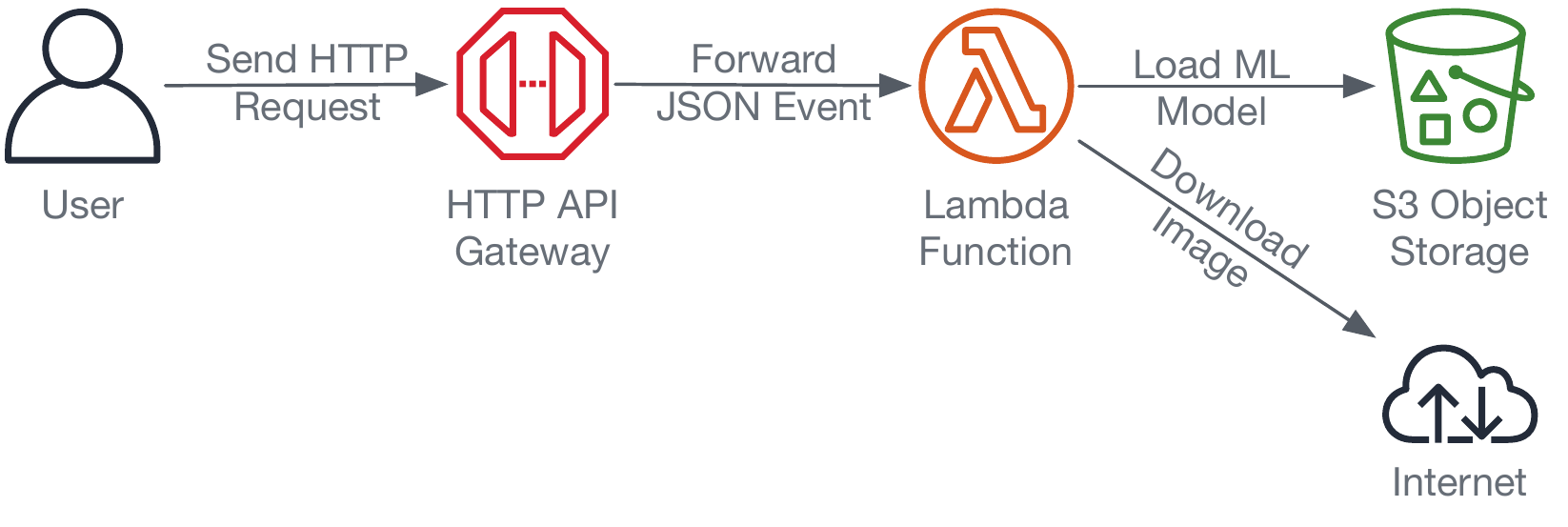}
    \caption{FaaS application for machine learning (ML) inference}\label{fig:ml_app}
\end{figure}

Figure~\ref{fig:ml_app} depicts the architecture of an AWS Lambda FaaS application that performs machine learning (ML) inferencing.
The diagram is based on the \code{mxnet-lambda} reference implementation\footurl{https://github.com/awslabs/mxnet-lambda} used in adjusted form by one study~[\ref{a16}] to benchmark ML inferencing.
The application predicts image labels for a user-provided image using a pre-trained deep learning model.
A user interacts with the application by sending an HTTP request to the HTTP API gateway, which transforms the incoming HTTP request into a cloud event and triggers an associated lambda function.
The API gateway serves as an example of a common function trigger.
However, lambda functions can also be triggered programmatically (e.g., via CLI or SDK), by other cloud events, such as file uploads (e.g., creation or modification of objects in S3), or various other trigger types.

Lambda functions implement the actual application logic, in our example application by loading the pre-trained ML model from S3, downloading the image from the user-provided URL from the internet, and then performing the inference computation within the lambda function.
Such lambda functions commonly make use of cloud services for data storage (e.g., object storage S3, document database DynamoDB), logging (e.g., CloudWatch monitoring), analytics (e.g., Amazon EMR including Hadoop, Spark, HBase, and other big data frameworks), machine learning (e.g., natural language translation with AWS Translate) and many more purposes.
The image download exemplifies other potential interactions with third-party services, such as REST APIs.
Finally, our interactive example application returns the address and geographical coordinates (i.e., the predicted image labels) to the user through the HTTP API gateway as an HTTP response.
In other non-interactive scenarios, lambda functions typically deliver their results to other cloud services, which might themselves trigger further actions or even other lambda functions as part of a workflow.

\section{Research Questions}\label{sec:rq}

In the context of studies on FaaS performance evaluation, our research questions address publication trends (\ref{rq:trends}), benchmarked platforms (\ref{rq:platforms}), evaluated performance characteristics (\ref{rq:characteristic}), used platform configurations (\ref{rq:config}), and reproducibility (\ref{rq:reproducibility}):

\begin{myrq}{rq:trends}
    \textbf{Publication Trends:} What are the publication trends related to FaaS performance evaluations?
\end{myrq}

This question helps us understand how active research on FaaS performance evaluation has been and gives us insights on publication types and academic venues.
This type of question is common to systematic mapping studies and has been studied in previous work for FaaS in general~\citep{yussupov:19} and for other domains~\citep{garousi:17}.

\begin{myrq}{rq:platforms}
    \textbf{Benchmarked Platforms:} Which FaaS platforms are commonly benchmarked?
\end{myrq}

This question intends to identify FaaS platforms that are particularly well-understood or under-researched.

\begin{myrq}{rq:characteristic}
    \textbf{Evaluated Performance Characteristics:} Which performance characteristics have been benchmarked for which FaaS platforms?
\end{myrq}

This question aims to characterize the landscape of existing work on FaaS performance to systematically map prior work and guide future research.

We divide the performance characteristics into the following sub-questions:
\begin{mysubrq}{rq:characteristic_types}
    \textbf{Evaluated Benchmark Types:} Are experiments typically using micro- or application-level benchmarks?
\end{mysubrq}
\begin{mysubrq}{rq:characteristic_micro}
    \textbf{Evaluated Micro-Benchmarks:} Which micro-benchmarks (e.g., CPU or IO benchmarks) are commonly evaluated?
\end{mysubrq}
\begin{mysubrq}{rq:characteristic_general}
    \textbf{Evaluated General Characteristics:} Which general performance characteristics (e.g., platform overhead / cold starts) are commonly evaluated?
\end{mysubrq}

\begin{myrq}{rq:config}
    \textbf{Used Platform Configurations:} Which platform configurations are commonly used?
\end{myrq}

This question targets the depth of the current understanding on FaaS performance.
We want to examine whether many studies conduct similar experiments or explore diverse configurations.

We divide the platform configurations into the following three sub-questions:
\begin{mysubrq}{rq:config_runtimes}
    \textbf{Used Language Runtimes:} Which language runtimes are commonly used?
\end{mysubrq}
\begin{mysubrq}{rq:config_triggers}
    \textbf{Used Function Triggers:} Which function trigger types are commonly used?
\end{mysubrq}
\begin{mysubrq}{rq:config_services}
    \textbf{Used External Services:} Which external services are commonly used?
\end{mysubrq}

\begin{myrq}{rq:reproducibility}
    \textbf{Reproducibility:} How reproducible are the reported experiments? %
\end{myrq}

This question addresses an inherently important quality of experimental designs by assessing how well the FaaS community follows existing guidelines on reproducible experimentation in cloud systems~\citep{papadopoulos:19}.

\section{Study Design}\label{sec:study}

This section describes the methodology of our Multivocal Literature Review (MLR) based on the guidelines from \citet{garousi:19}.
We first summarize the overall process, then detail the strategies for search, selection, and data extraction and synthesis, followed by a discussion of threats to validity.

\subsection{MLR Process Overview}\label{sec:process}
The MLR process is divided into a part for academic and grey literature.
We classify peer-reviewed papers (e.g., papers published in journals, conferences, workshops) as academic literature (i.e., white literature) and other studies (e.g., preprints of unpublished papers, student theses, blog posts) as grey literature.

\begin{figure}
    \centering
    \includegraphics[width=0.4\textwidth]{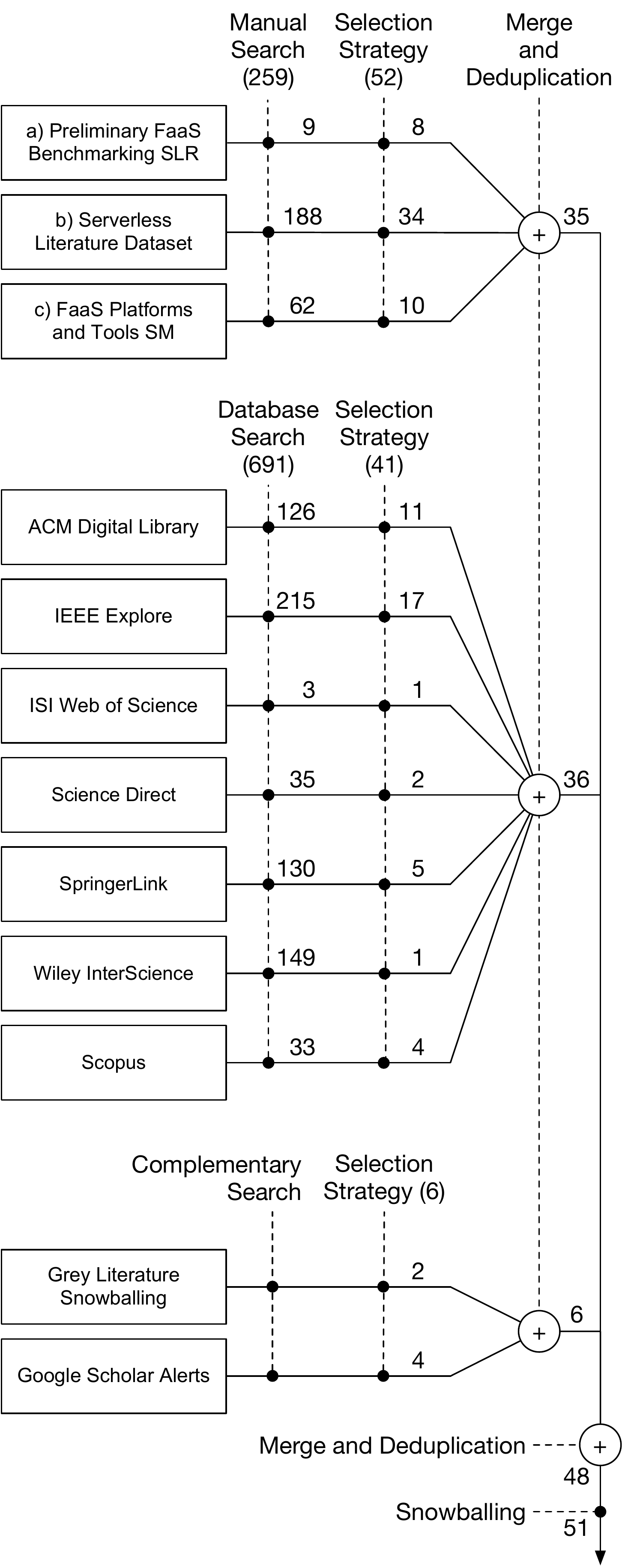}
    \caption{Process for systematic literature review}\label{fig:mlr-process-academic}
\end{figure}

The search process and source selection for academic literature follow a conventional systematic literature review (SLR) process~\citep{kitchenham:07}.
Figure~\ref{fig:mlr-process-academic} summarizes this multi-stage process originating from three different search sources and annotates the number of studies after each stage.

\begin{figure}
    \centering
    \includegraphics[width=0.47\textwidth]{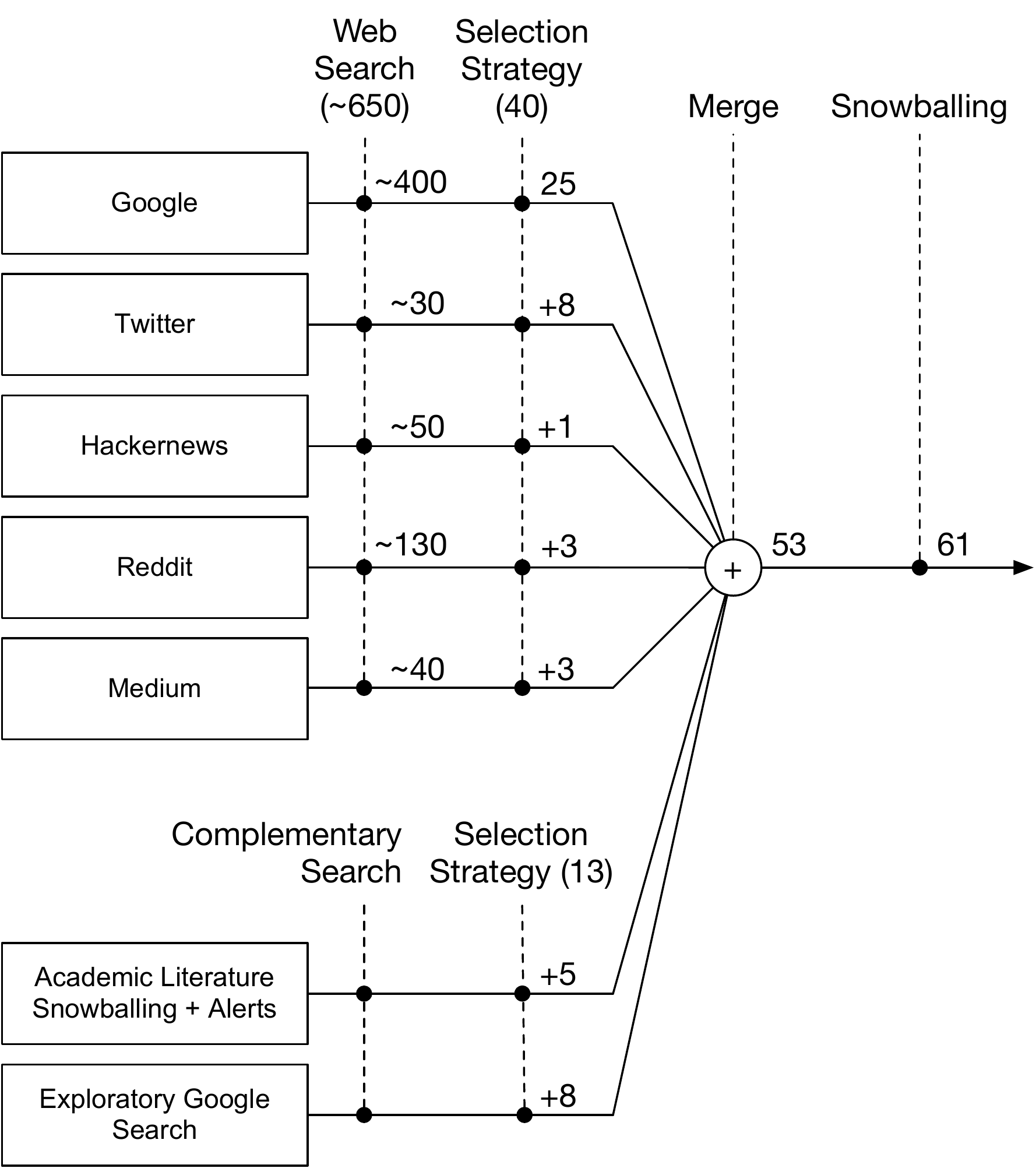}
    \caption{Process for grey literature review}\label{fig:mlr-process-grey}
\end{figure}

The process for grey literature studies is summarized in Figure~\ref{fig:mlr-process-grey} with sources originating prevalently from web search.
Notice that the number of relevant studies are already deduplicated, meaning that we found 25 relevant studies through Google search and the additional +8 studies from Twitter search only include new, non-duplicate studies.
A key motivation for the inclusion of grey literature is the strong industrial interest in FaaS performance and the goal to identify potential mismatches between academic and industrial perspectives.

\subsection{Search Strategies}
We first describe manual and database search for academic publications, then highlight the adjustments for web search, and finally discuss how alert-based search and snowballing complement the classic search strategies.
For manual, database, and web search, we were able to perform an exhaustive search by applying the selection strategy to all initial search results.

\subsubsection{Manual Search for Academic Literature} %
We use manual search to establish an initial seed of relevant sources to refine the database search query and to complement database search results with sources from third-party literature collections.
We screen the following sources for potentially relevant studies:
\begin{itemize}
    \item[a)] Studies from the preliminary results of an SLR targeting benchmarking of FaaS platforms~\citep{kuhlenkamp:18}:
    Their references from Table~1 are all relevant for our MLR but limited to 9 FaaS benchmarking studies, from which we removed one due to duplication (a journal extension covering more experiments than the initial extended abstract).
    \item[b)] Studies from the \emph{Serverless Literature Dataset}~\citep{spillner:19a} (Version 0.4) listing 188 serverless-related papers published between 2016 and 2019: %
    This extensive list of references covers many different serverless topics and therefore only a small subset of studies are potentially relevant for our MLR.
    \item[c)] Studies from a systematic mapping (SM) study on engineering FaaS platforms and tools~\citep{yussupov:19}:
    Their 62 selected publications focus on novel approaches and thus explicitly exclude benchmarking studies \enquote{without proposing any modifications}~\citep{yussupov:19}.
    We still identify a total of 10 relevant studies for our MLR in their categories related to benchmarking and performance and by screening their complete references.
\end{itemize}

\subsubsection{Database Search for Academic Literature}
Following standard SLR guidelines~\citep{kitchenham:07}, we define a search string to query common digital libraries for potentially relevant papers.
We make use of logical OR operators to consider alternative terms given the lack of terminology standardization in the field of serverless computing.
Within the area of serverless computing (\ie \code{(serverless OR faas)}), our search string targets performance-related (i.e., \code{performance OR benchmark}) empirical (\ie \code{experiment}) research.
We refine the search string based on the insights from manual search, as suggested by \citet{zhang:11a}, by adding the additional keyword \code{lambda} (appeared in all full texts) but omitting double quotes for exact matching.
Our final search string is defined as follows:\\
\code{(serverless OR faas) AND (performance OR benchmark)\\
    AND experiment AND lambda
}

We apply the search string to 7 common digital libraries, namely
ACM Digital Library, IEEE Explore, ISI Web of Science, Science Direct, Springer Link, Wiley InterScience, and Scopus.
The libraries are configured in their advanced query modes (if available) to search within full texts and metadata fields for maximal search coverage.
The exact search query for each library can be found in the online appendix\footurl{\appendixurl} including direct links and instructions for reproducing the search.
The search was performed in October 2019 and all raw search results are exported into the bibtex format.

\subsubsection{Web Search for Grey Literature}

For querying grey literature, we modified our original search string to account for less formal language in online articles.
We replicate our academic search for one Google query but omit the terms \emph{experiment} and \emph{lambda} for all remaining queries using the following simplified search string:
\code{(serverless OR faas) AND (performance OR benchmark)}\\
We apply the search string to 5 search engines, namely Google Search, Twitter Search, Hacker News Algolia Search, Reddit Search, and Medium Search.
These engines (except Google Search) lack support for logical \emph{OR} expressions.
Therefore, we compose and combine four logically equivalent subqueries equivalent to the defined search string.
Most searches were performed in Dec 2019 and, for replicability, we save the output of every search query as PDF and HTML files.
Notice that the numbers of initial search results for web search are rough estimates due to the nature of web search engines.
We refer to our replication package for technical details~\citep{scheuner:20-jss-dataset}.

\subsubsection{Complementary Search}
Our previous search strategies often missed recent literature because manual search heavily relies on previous work and database search might suffer from outdated query indices or omit academic literature in press (i.e., accepted but not yet published).
To discover recently published literature, we configured Google Scholar web-alerts\footurl{https://scholar.google.com/intl/en/scholar/help.html\#alerts} for the broad search term \code{serverless} and the more specific search term \code{serverless benchmark} over a period of 5 months (2019-10 till 2020-02) and screened hundreds of articles for potential relevance.
Alert-based search discovered 6 relevant preprints (e.g., from arXiv.org) for which we explicitly checked whether they were accepted manuscripts (4 academic literature) or unpublished preprints (2 grey literature).
Hence, complementary search for academic literature also contributed relevant studies to grey literature through alerts and snowballing, as well as vice versa.
For grey literature, we spotted further relevant studies through exploratory Google search following looser adaptations of the search terms in particular contexts (e.g., related to a benchmarking tool).

\subsubsection{Snowballing}

After applying the selection criteria, we perform snowballing for academic and grey literature.
For academic literature, we apply backward snowballing by screening their reference lists and forward snowballing by querying citations to relevant papers through Google Scholar. %
For grey literature, we prevalently apply backward snowballing by following outgoing links and occasionally (particularly for popular and highly relevant sources) apply forward snowballing by querying incoming for links through a backlink checker\footurl{https://ahrefs.com/backlink-checker}.

\subsection{Selection Strategy}\label{sec:selection}

Following established SLR study guidelines~\citep{kitchenham:07}, we define the following inclusion (I) and exclusion (E) criteria for our study:
\begin{itemize}
    \item[I1] Studies performed at least one performance-related experiment (i.e., excluding purely theoretical works, simulations, and works where a performance experiment was only mentioned as a sidenote) with a real FaaS environment as System-Under-Test (SUT). The FaaS environment can be fully managed or self-hosted.
    \item[I2] Studies presented empirical results of at least one performance metric.
    \item[I3] Studies published after Jan 1st 2015, as the first FaaS offering (AWS Lambda) was officially released for production use on April 9, 2015\footurl{https://docs.aws.amazon.com/lambda/latest/dg/lambda-releases.html}.
    \item[E1] Studies written in any other language than English
    \item[E2] Secondary or tertiary studies (e.g., SLRs, surveys)
    \item[E3] Re-posted or republished content (e.g., sponsored re-post, conference paper with a journal extension)
\end{itemize}

As suggested by \citet{wohlin:12}, we only consider the most complete study as relevant primary study in cases of partial republication, for instance in the case of a journal extension of a conference paper.
The two authors classified each potentially relevant study either as \emph{relevant}, \emph{uncertain} (with an indication whether rather relevant or not), or \emph{not relevant}.
All studies classified as \emph{uncertain} were examined again and the rationale for the final decision was documented following the selection strategy presented above.
If the title, keywords, and abstract were insufficient for obviously excluding a study, we read the full text of the study to take a final decision as practiced for all included studies.

\subsection{Data Extraction and Synthesis}\label{sec:extract}

Guided by the research questions, we extract the corresponding information based on a structured review sheet.

\paragraph{Publication Trends (\ref{rq:trends})}
To capture how many studies of which type are published, we extract the following metadata: (i) the publication date (ii) the venue type for academic literature (i.e., journal, conference, workshop, doctoral symposium) and grey literature (i.e., preprint, thesis, blog post) (iii) the name of the venue (e.g., IEEE CLOUD, USENIX ATC), and a ranking of the venue (i.e., A*, A, B, C, W for workshop, unranked).
The venue ranking follows the CORE ranking for conferences (CORE2018\footurl{http://portal.core.edu.au/conf-ranks/}) and journals (ERA2010\footurl{http://portal.core.edu.au/jnl-ranks/}).

\paragraph{Benchmarked Platforms (\ref{rq:platforms})}
To assess which offerings are particularly well-understood or insufficiently researched, we extract the names of all FaaS platforms that are empirically investigated in a study.

\paragraph{Evaluated Performance Characteristics (\ref{rq:characteristic})}
To understand which performance characteristics have been benchmarked, we distinguish between micro- and application-benchmarks, collect a list of micro-benchmarks (e.g., CPU speed, network performance), and capture more general performance characteristics (e.g., use of concurrent execution, inspection of infrastructure).
We start with an initial list of characteristics and iteratively add popular characteristics from an open \emph{Others} field.

\paragraph{Used Platform Configurations (\ref{rq:config})}
To describe which platform configurations have been evaluated, we extract the list of used language runtimes (\ref{rq:config_runtimes}), function triggers (\ref{rq:config_triggers}), and external services (\ref{rq:config_services}).
We generalize vendor-specific services to cross-platform terminology (e.g., AWS S3 was generalized to cloud storage).

\paragraph{Reproducibility (\ref{rq:reproducibility})}
To review the potential regarding reproducibility, we follow existing guidelines on experimentation in cloud systems~\citep{papadopoulos:19}.
The authors propose eight fundamental methodological principles on how to measure and report performance in the cloud and conduct an SLR to analyze the current practice concerning these principles covering top venues in the general field of cloud experimentation.
As part of our work, we replicate their survey study in the more specific field of FaaS experimentation.
We largely follow the same study protocol by classifying for each principle whether it is fully met (\emph{yes}), partially present (\emph{partial}) but not comprehensively following all criteria, or not present (\emph{no}).
Additionally, we collect some more fine-grained data for certain principles.
For example, we distinguish between dataset availability and benchmark code availability for P4 (open access artifact) because we consider public datasets to be essential for replicating (statistical) analyses and public benchmark code is practically essential for reproducing the empirical experiment.
For P3 (experimental setup description), we additionally capture whether a study describes the time of experiment (i.e., dates when the experiment was conducted), cloud provider region (i.e., location of data center), and function size (i.e., used memory configurations).

\subsection{Threats to Validity}\label{sec:threats}

We discuss potential threats to validity and mitigation strategies for selection bias, data extraction and internal validity, replicability of the study, and external validity.

\paragraph{Selection Bias}
The representativeness of our selected studies is arguably one of the main threats to this study.
We used a multi-stage process (see Section~\ref{sec:process}) with sources originating from different search strategies.
Initial manual search based on existing academic literature collections allowed us to fine-tune the query string for database searches against 7 well-established electronic research databases.
We optimize our search string for more informal grey literature and query 5 search engines specializing in general-purpose search, social search, and developer-community search.
Additionally, our complementary search strategies aim to discover studies that were recently published, found in the other context (i.e., academic vs grey), or spotted through more exploratory search (e.g., a looser adaptation of search terms).

\paragraph{Data Extraction and Internal Validity}
Tedious manual data extraction could potentially lead to inaccuracies in extracted data.
To mitigate this threat, we define our MLR process based on well-established guidelines for SLR~\citep{kitchenham:07} and MLR~\citep{garousi:19} studies, methodologically related publications~\citep{garousi:17}, and topically relevant publications~\citep{yussupov:19,kuhlenkamp:18}.
Further, we set up a structured review sheet with practical classification guidelines and further documentation, which was incrementally refined (e.g., with advice on classifying borderline cases).
We implemented traceability through over 700 additional comments, at least for all borderline cases.
The data extraction process was conducted by both authors, with the first author as the main data extractor and the second author focusing on discussing and verifying borderline cases.
We also repeatedly went over all sources to verify certain data (e.g., based on refined classification scheme) and collect more details (e.g., individual aspects of more vague P3 on experimental setup description).
For the reproducibility part (see \ref{rq:reproducibility}), we refer to the statistical evaluation on inter-reviewer agreement in the original study~\citep{papadopoulos:19}, which achieved very high agreement.

\paragraph{Replicability of the Study}
We publish a replication package~\citep{scheuner:20-jss-dataset} to foster verification and replication of our MLR study.
Our package includes all search queries with direct links and step-by-step instructions on how to replicate the exact same queries, query results in machine-readable (BibTeX/HTML) and human-readable (PDF) formats, a structured review sheet containing all extracted data and over 700 comments with guidance, decision rationales, and extra information, and code to reproduce all figures in our study. %
The latest version of the replication package and further documentation is also available online\footurl{\appendixurl}.

\paragraph{External Validity}
Our study is designed to systematically cover the field of FaaS performance benchmarking for peer-reviewed academic white literature and unpublished grey literature including preprints, theses, and articles on the internet.
However, we cannot claim generalizability to all academic or white literature as we might have missed some studies with our search strategies.
The inclusion of grey literature aims to address an industrial perspective but is limited to published and indexed content freely available and discoverable on the internet (e.g., excluding paywall articles or internal corporate feasibility studies).

\section{Study Results and Discussion}\label{sec:results}

This section presents and discusses the main outcomes of our MLR study guided by our research questions stated in Section~\ref{sec:rq}.
The results are based on the extracted and synthesized (according to Section~\ref{sec:extract}) survey data from 112 selected (according to Section~\ref{sec:selection}) primary studies including 51 academic publications and 61 grey literature sources.
For each research question, we briefly describe context, motivation, and methodology, followed by relevant results and their subsequent discussion.

\subsection{Publication Trends (\ref{rq:trends})}

\paragraph{Description}
We describe the publication trends on FaaS performance evaluations by summarizing the publication statistics over years and venue types, the venue rankings for academic literature, and the most popular publication venues.
The venue ranking follows the CORE ranking for conferences (CORE2018) and journals (ERA2010).

\begin{figure}[htb]
    \centering
    \includegraphics[width=0.46\textwidth]{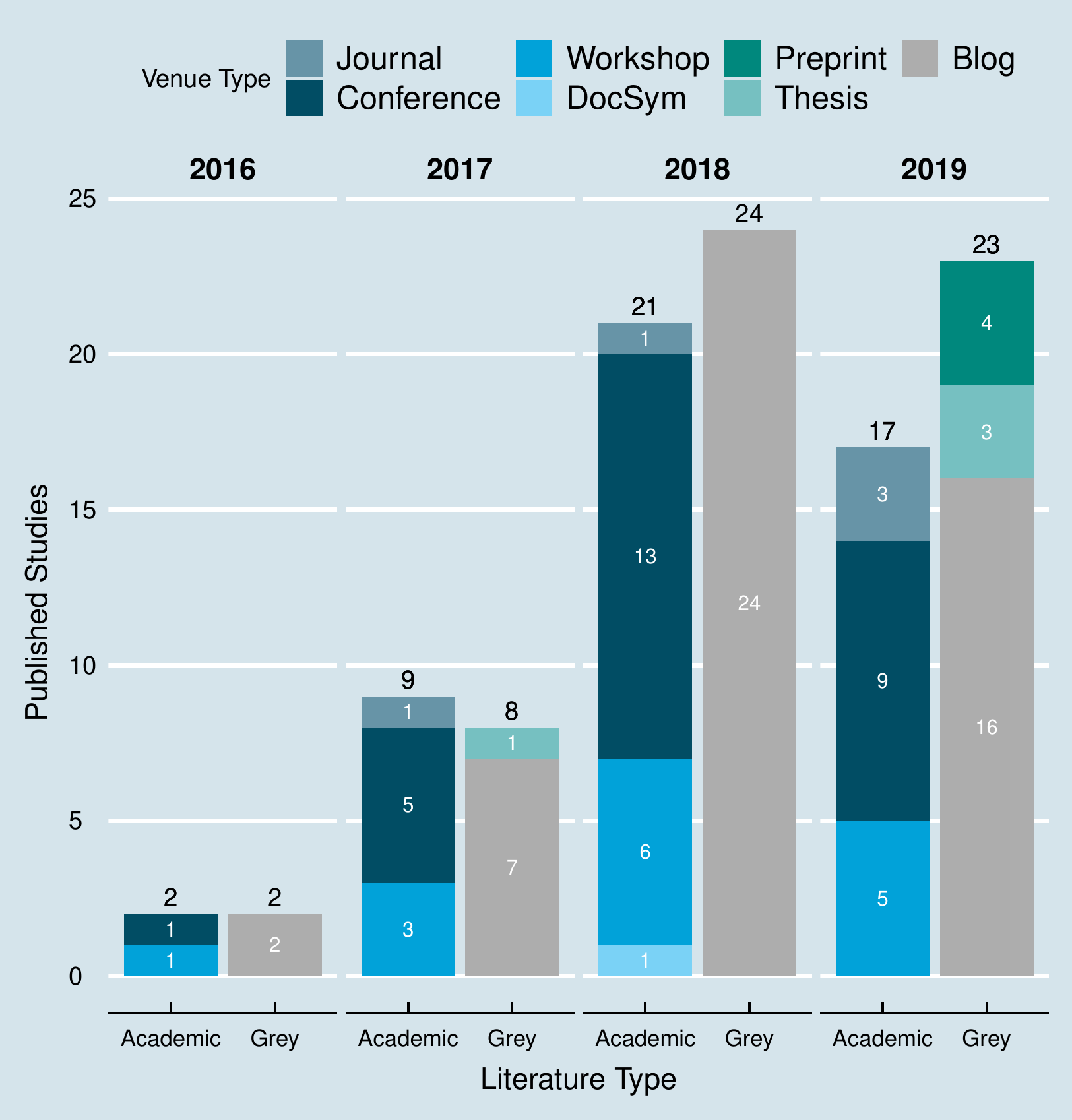}
    \caption{Distribution of academic (N=51) and grey (N=61) literature studies over years and venue type}\label{fig:trends}
\end{figure}

\paragraph{Results}
Figure~\ref{fig:trends} shows the distribution of published studies for academic and grey literature over years and venue types.
We observe a growing interest for both types of literature, with early studies appearing in mid 2016~[\ref{a15}, \ref{a35}], followed by a drastic increase in 2017, and a surge of new studies in 2018.
The year 2019 indicates a minor decrease in overall publication activity but covers more diverse publication types.
Notice that the initial searches were performed in October 2019 for academic literature and December 2019 for grey literature and therefore cover 2019 only partially, also considering the indexing delay.
Further, Figure~\ref{fig:trends} omits 3 blog posts with unspecified publication dates and also the just-started year 2020 consisting of 2 academic conference publications and 1 grey literature preprint.

\begin{figure}[htb]
    \centering
    \includegraphics[width=0.46\textwidth]{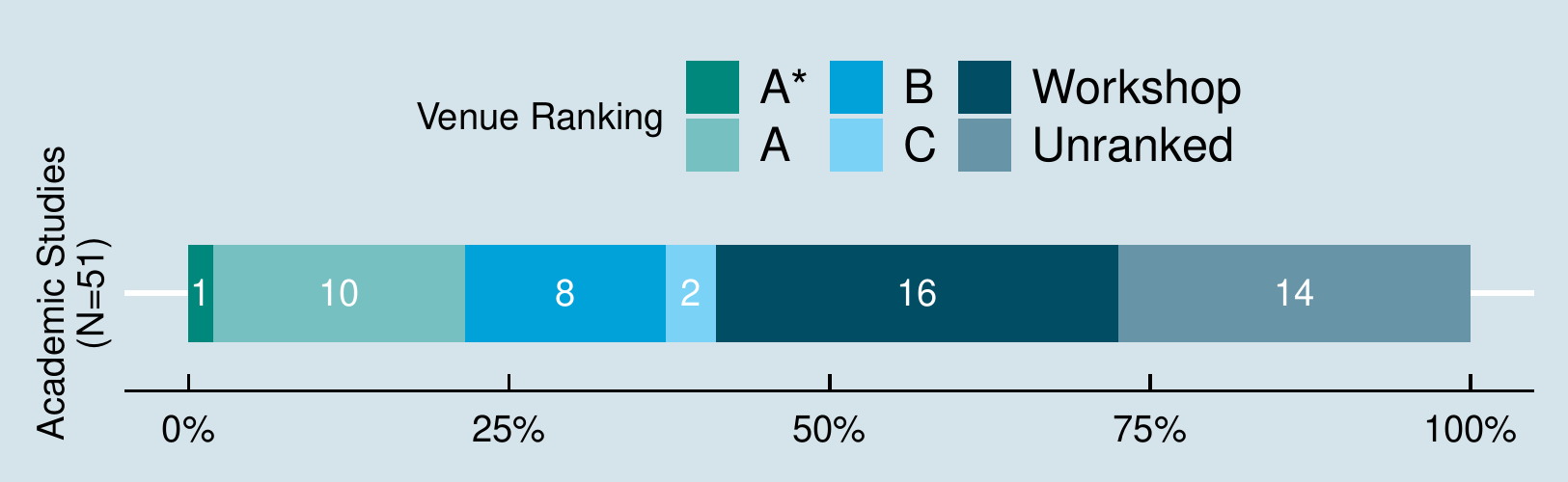}
    \caption{Academic venue rankings (N=51)}\label{fig:rankings}
\end{figure}

Figure~\ref{fig:rankings} summarizes the venue ranking across all academic studies.
A good share of studies is published in top-ranked venues, surprisingly few studies appear in C-ranked venues, and the majority of studies are published in workshops or other unranked venues.

Table~\ref{tab:venues} lists the 10 most popular publication venues where at least two studies have been published.
The most popular venue with 7 publications is the International Workshop on Serverless Computing (WoSC), founded in 2017 and held two times per year in 2017 and 2018.
\begin{table}[htb]
    \centering
    \scriptsize
    \caption{\label{tab:venues}List of 10 most popular publication venues.}
    \begin{tabular}{@{}llp{6.5cm}@{}}
        \toprule
        \textbf{N} & \textbf{Acronym} & \textbf{Venue}                                                  \\ \midrule
        7          & WoSC             & International Workshops on Serverless Computing                 \\
        5          & ATC              & USENIX Annual Technical Conference                              \\
        3          & CLOUD            & IEEE International Conference on Cloud Computing                \\
        2          & SAC              & ACM/SIGAPP Symposium On Applied Computing                       \\
        2          & FGCS             & Future Generation Computer Systems                              \\
        2          & ICSOC            & International Conference on Service Oriented Computing          \\
        2          & BigData          & IEEE International Conference on Big Data                       \\
        2          & IC2E             & IEEE International Conference on Cloud Engineering              \\
        2          & HotCloud         & USENIX Workshop on Hot Topics in Cloud Computing                \\
        2          & NSDI             & USENIX Symposium on Networked Systems Design and Implementation \\ \bottomrule
    \end{tabular}
\end{table}

\paragraph{Discussion}
Our results are generally in line with the related systematic mapping study from \citet{yussupov:19}.
However, we see a stronger emphasis on workshop publications, which appears plausible for a more narrow topic of investigation.
Additionally, our work indicates that grey literature follows a similar but possibly more pronounced hype trend with blog posts spiking in 2018 and declining stronger in 2019 than cumulative academic literature.

Related to academic venue rankings, we interpret the relative over-representation of top-ranked publications (in comparison to relatively few full papers in C-ranked venues) as a positive sign for this young field of research.
The strong representation of workshop papers, particularly at WoSC, is plausible for a relatively narrow topic in a young line of research.

\subsection{Benchmarked Platforms (\ref{rq:platforms})}

\paragraph{Description}
The two main types of FaaS platforms are hosted platforms and platforms intended to be installed in a private cloud.
Hosted platforms are fully managed by a cloud provider and often referred to as FaaS providers.
All major public cloud providers offer FaaS platforms, including AWS Lambda, Microsoft Azure Functions, Google Cloud Functions, and IBM Cloud Functions.
Installable platforms are provided as open source software and can be self-hosted in on-premise deployments.
Prominent open source platforms include Apache OpenWhisk, Fission, Knative, or OpenFaaS.
Self-hosting requires extra setup, configuration, and maintenance efforts, but allows for full control and inspection during experimentation.
Dozens more hosted services\footurl{https://landscape.cncf.io/format=serverless} and many more FaaS development frameworks\footurl{https://github.com/anaibol/awesome-serverless\#frameworks}
and installable platforms have emerged in this fast-growing market.

\begin{figure*}
    \centering
    \includegraphics[width=\textwidth]{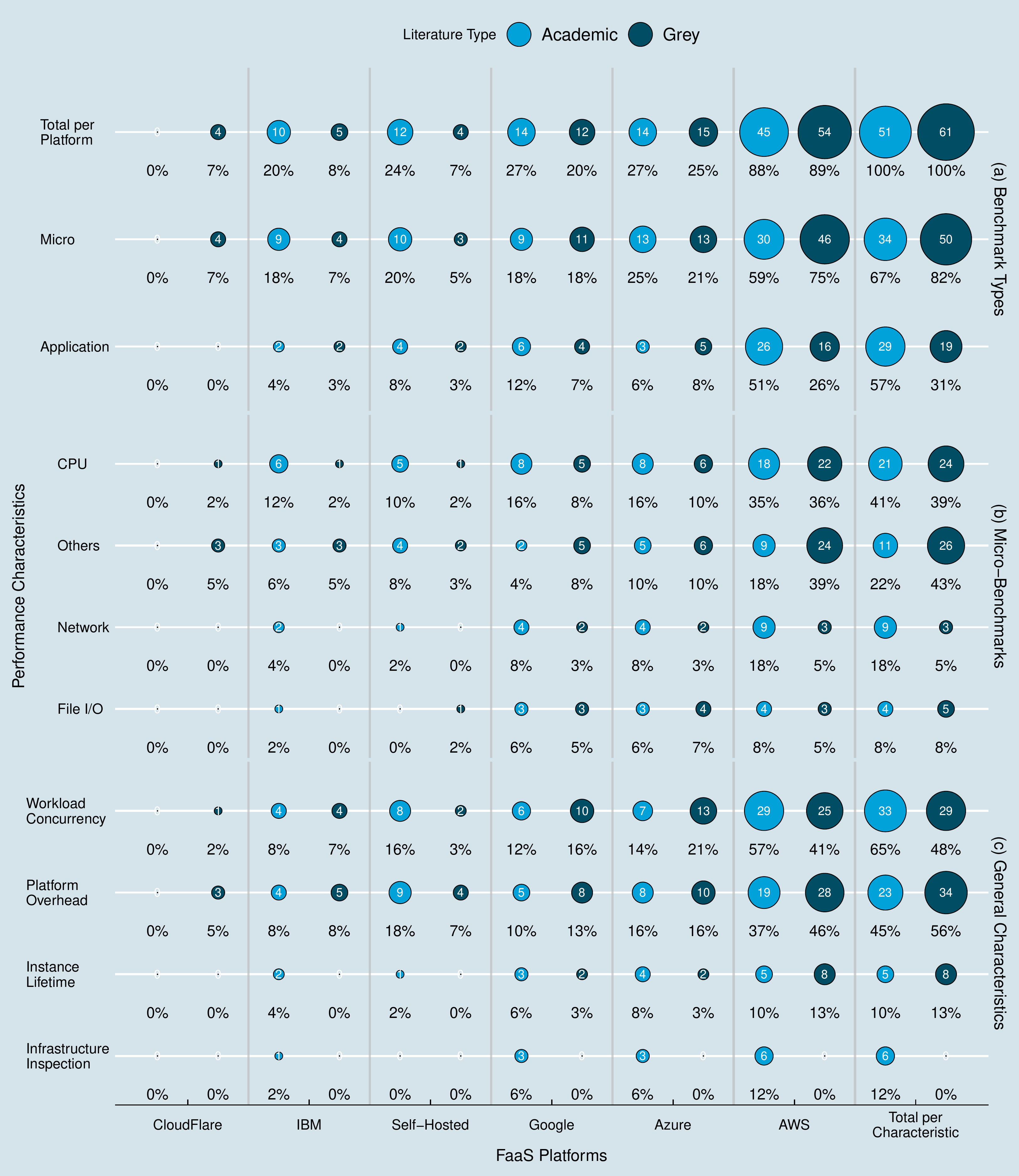}
    \caption{Evaluated performance characteristics per FaaS platform}\label{fig:characteristics}
\end{figure*}

\paragraph{Results}
The first row of the bubbleplot in Figure~\ref{fig:characteristics}a summarizes the total number of performance evaluation experiments in absolute frequency counts for the 5 most popular hosted FaaS platforms in our study.
Self-hosted platforms are only depicted in aggregation due to their low prevalence in literature.
The x-axis is ordered by cumulative platform frequency, where AWS Lambda leads with a total of 99 studies divided into 45 academic and 54 grey literature studies.
Thus, 88\% of all our selected studies perform experiments on AWS Lambda, followed by Azure (26\%), Google (23\%), self-hosted platforms (14\%), IBM (13\%), and CloudFlare (4\%).
For hosted platforms, we omit Lambda@Edge\footurl{https://aws.amazon.com/lambda/edge/} (3) and Binaris\footnote{Binaris (\url{https://binaris.com/}) was renamed to reshuffle in Oct~2019} (1) because Lambda@Edge is covered in the same experiments as CloudFlare and Binaris only occurred once.
Within self-hosted platforms, academic literature mostly focuses on OpenWhisk (70\%), whereas grey literature covers other platforms, such as Fission, Fn, or OpenFaaS.

\paragraph{Discussion}\label{sec:discussion_platforms}
In comparison to other surveys, our overall results for percentage by provider closely ($\pm$5\%) match the self-reported experience per cloud provider in a 2018 FaaS survey (N$=$182)~\citep{leitner:19}.
Our results are also reasonably close ($\pm$5\% except for AWS +13\%) to self-reported use in organizations in a 2019 O’Reilly survey on serverless architecture adoption (N>1500)\footnote{Figure 12 in \url{https://www.oreilly.com/radar/oreilly-serverless-survey-2019-concerns-what-works-and-what-to-expect/}}.
Initial results of the latest 2020 survey (N>120 for the first day)\footnote{Question 14 in \url{https://www.nuweba.com/blog/serverless-community-survey-2020-results}} indicate similar results for FaaS products currently in use, with a very close match for AWS (1\% deviation).
However, this survey shows even lower numbers (up to \SI{-12}{\percent}) for other providers.

Hence, our results show that AWS is currently overstudied in absolute numbers (by a factor of 3x).
However, the strong emphasis on AWS appears to be justified in relative numbers given the industrial importance of AWS in this domain.
We observe that the attention by literature type is appropriately balanced (less than $\pm10\%$) for most platforms, except for proportionally higher academic coverage by IBM (+12\%) and self-hosted platforms (+17\%).
IBM appears to be over-represented in academic studies, potentially motivated by the ability to compare a provider-hosted platform with its underlying open source platform Apache OpenWhisk in a self-hosted setup (e.g., \ref{a01}, \ref{a03}).
In contrast to hosted platforms, self-hosted platforms allow for full control of the experimental setup (e.g., \ref{a28}) and detailed performance and resource profiling (e.g., \citet{shahrad:19}) but raise other challenges regarding fair comparability and configuration.

\subsection{Evaluated Performance Characteristics (\ref{rq:characteristic})}

To answer \ref{rq:characteristic}, the facetted bubbleplot in Figure~\ref{fig:characteristics} combines performance characteristics for a) benchmark types b) micro-benchmarks, and c) general characteristics across FaaS platforms.
All these plots can be interpreted as a heatmap ranging from few studies in the bottom-left corner to many studies in the top-right corner for a given characteristic-platform combination.
We provide relative percentages against the total number per literature type (i.e., $N_{academic}=51$ vs $N_{grey}=61$) because the absolute numbers are not directly comparable.

\subsubsection{Evaluated Benchmark Types (\ref{rq:characteristic_types})}

\begin{figure}[htb]
    \centering
    \includegraphics[width=0.4\textwidth]{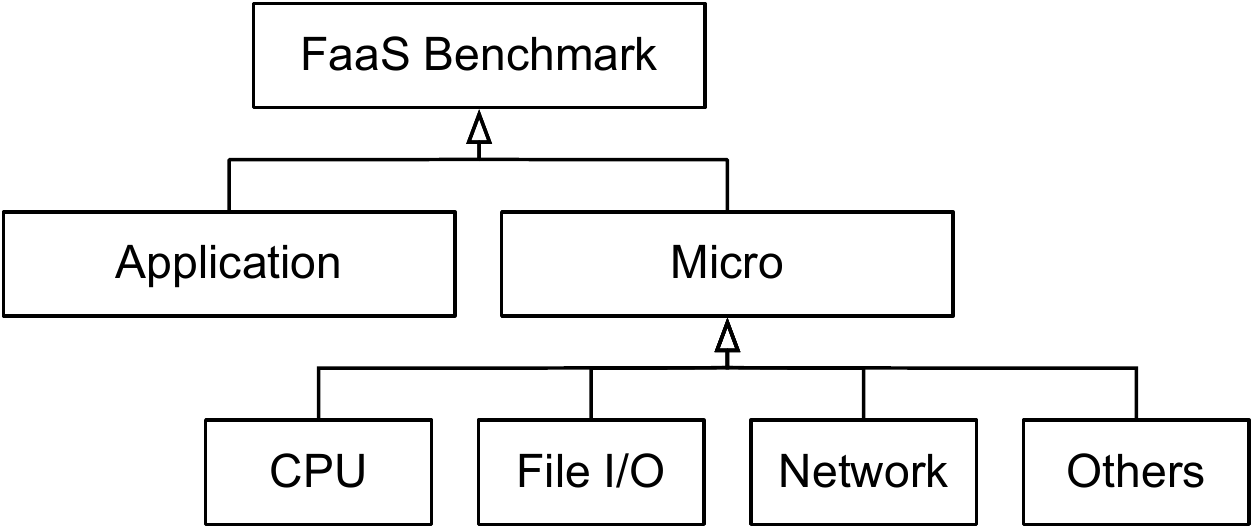}
    \caption{FaaS benchmark taxonomy}\label{fig:benchmark_taxonomy}
\end{figure}

\paragraph{Description}
We distinguish between narrow micro- and holistic application-benchmarks as introduced in Section~\ref{sec:bg}.
Figure~\ref{fig:benchmark_taxonomy} summarizes our FaaS benchmark taxonomy including the most common micro-benchmarks derived through open coding as described in Section~\ref{sec:extract} and reflected in Figure~\ref{fig:characteristics}.

\paragraph{Results}
Figure~\ref{fig:characteristics}a summarizes which high-level types of benchmarks are used across which FaaS platforms.
The rightmost \emph{Total per Characteristic} column indicates that micro-benchmarks are the most common benchmark type, used by 75\% of all selected studies (84/112).
Interestingly, we observe a particularly strong emphasis on micro-benchmarks in grey literature (50 studies, or 82\%). However, also two-thirds of the selected academic literature conduct studies using micro-benchmarks.
Application-level benchmarks are used by 48 (43\%) of all selected studies and are more prevalent among academic literature with 29 (57\%) studies compared to grey literature with only 19 (31\%) studies.
Further, 12 (24\%) academic studies combine micro-benchmarks and application-benchmarks, which can be derived by the difference (i.e., 12) between the total number of academic literature studies (51) and the sum of total micro-benchmarks and application-benchmarks ($34+29=63$).
For grey literature, only 8 (13\%) studies combine the two benchmark types and thus the vast majority of studies (87\%) uses micro- or application-benchmarks in isolation.
Finally, micro-benchmarks are more commonly used across different providers, whereas application-benchmarks are prevalently (\textgreater 84\%) used for benchmarking AWS.

\paragraph{Discussion}
Micro-benchmarks are the most common benchmark type across both literature types but academic literature uses application-benchmarks more frequently (+26\%) than grey literature.
We expected application-benchmarks to be under-represented but were surprized to see relatively many academic studies using application-benchmarks. %
Closer investigation revealed that many of these academic studies demonstrate or evaluate their proposed prototypes on a single FaaS platform (e.g., \enquote{MapReduce on AWS Lambda} [\ref{a12}]) focusing on thorough insights and leaving cross-platform comparisons for future work.
While we agree that such studies on a single FaaS platform can be great demonstrators of ideas and capabilities, the general usefulness of application-benchmarks evaluated in isolation on a single platform is limited, as
the inability to relate results from such work against any baseline or other reference platform makes it hard to derive meaningful conclusions.
This threat is particularly relevant for hosted platforms, as the performance observed from such experiments depends strongly on the underlying hard- and software infrastructure.
Therefore, we argue that reproducibility (see \ref{rq:reproducibility}) is particularly important for this type of study.

Some studies clearly intend to conduct end-to-end (i.e., application-level) measurements, however, their applications and workloads are insufficiently described such that it is unclear what exactly they do.
This unclarity is reinforced by the tendency of application-benchmarks to remain closed source, with only 35\% of the studies publishing at least partial benchmark code compared to 50\% overall.

\subsubsection{Evaluated Micro-Benchmarks (\ref{rq:characteristic_micro})}

\paragraph{Description}
We cover micro-benchmarks targeting CPU, file I/O, and network performance.
The \emph{Others} category summarizes other types of micro-benchmarks such as cold start evaluations (i.e., platform overhead).
Notice that we grouped platform overhead as general performance characteristics because some studies alternatively use application-benchmarks with detailed tracing.

\paragraph{Results}
Figure~\ref{fig:characteristics}b summarizes which micro-benchmark performance characteristics are used across which FaaS platforms.
The rightmost \emph{Total per Characteristic} column shows that CPU is by far the most evaluated micro-benchmark characteristic, used by 40\% of all studies.
Network and file I/O performance are less common for academic literature studies, and even more rare in grey literature.
These two less common characteristics are all evaluated on the AWS platform (except for two file-I/O grey literature studies) but practically uncovered on self-hosted platforms (only two studies overall).
The \emph{Others} category mainly consists of platform overhead and workload concurrency evaluated through micro-benchmarks.
While many studies evaluate different memory configurations or monitor memory usage, we have not seen studies evaluating memory performance (e.g., bandwidth, latency) itself.

\paragraph{Discussion}\label{sec:discussion_micro}

Academic literature tends to focus more on traditional performance characteristics, such as CPU, network, file I/O, in contrast to grey literature focusing more on other FaaS-specific characteristics, such as cold starts, concurrency, and trigger comparisons.

Our results suggest that CPU performance is an overstudied performance characteristic among FaaS micro-bench\-marks.
Many studies confirm that CPU processing speed scales proportionally to the amount of allocated memory (i.e., configured function size) for AWS~[\ref{a08}, \ref{a49}, \ref{a03}, \ref{g10}, \ref{g43}] and Google~[\ref{a08}, \ref{a49}, \ref{a03}, \ref{g10}].
This empirically validated behavior is also explicitly stated in the documentation of the providers.
For instance, the AWS Lambda documentation states that \enquote{Lambda allocates CPU power linearly in proportion to the amount of memory configured.}\footurl{https://docs.aws.amazon.com/lambda/latest/dg/configuration-console.html}.
The Google Cloud Functions documentation also used to mention proportional scaling explicitly.
A few of these studies~[\ref{a08}, \ref{a03}, \ref{g02}] also cover Azure and IBM and conclude that these platforms assign the same computational power for all functions.
Notice that Azure does not expose an explicit memory size configuration option as common for the other providers, but rather determines available memory sizes based on a customer's service subscription plan\footurl{https://docs.microsoft.com/en-us/azure/azure-functions/functions-scale\#service-limits}.

\subsubsection{Evaluated General Characteristics (\ref{rq:characteristic_general})}

\paragraph{Description}
We cover four general performance characteristics, namely platform overhead, workload concurrency, instance lifetime, and infrastructure inspection.
These general characteristics are orthogonal to previously discussed characteristics, and can be measured using either micro- or application-level benchmarks.
Platform overhead (e.g., provisioning of new function instances) mainly focuses on startup latency and in particular on quantifying the latency of cold starts. %
Workload concurrency refers to workloads that issue parallel requests, or to benchmarks evaluating platform elasticity  or scaling behavior (e.g., [\ref{a26}, \ref{a36}]).
Instance lifetime or infrastructure retention attempts to re-engineer the provider policy on how long function instances are kept alive until they get recycled and trigger a cold start upon a new function invocation.
Infrastructure inspection aims to re-engineer underlying hardware characteristics (e.g., CPU model to detect hardware heterogeneity) or instance placement policy (e.g., instance identifier and IP address to detect co-residency on the same VM/container~[\ref{a49}]).

\paragraph{Results}
Figure~\ref{fig:characteristics}c summarizes which general performance characteristics are benchmarked across which FaaS platforms.
Workload concurrency is a commonly studied characteristic, but more so in academic literature (65\%) than in grey literature (48\%).
On the other hand, grey literature seems to focus more on platform overhead (56\%) than academic literature (45\%).
Infrastructure inspection is exclusively analyzed in academic literature studies (12\%).
Note that this line of inquiry does not make sense for self-hosted platforms, and hence is not studied in this context.
Finally, the data from the \emph{Others} column shows that there is currently a lack of cross-platform comparisons of function triggers.

\paragraph{Discussion}
General performance characteristics focus on particularly relevant aspects of FaaS and only a few studies aim towards reverse-engineering hosted platforms.
Elasticity and automatic scalability have been identified as the most significant advantage of using FaaS in a previous survey~\citep{leitner:19}, which justifies the widespread evaluation of concurrency behavior.
Given the importance of this characteristic, we argue that concurrent workloads should be an inherent part of all FaaS performance evaluations going forward (going beyond the 50\% of studies observed in our corpus).
Container start-up latency has been identified as one of the major challenges for using FaaS services in prior work~\citep{leitner:19}, receives comparably even higher attention from grey literature (+11\%), and thus motivates a large body of work related to quantifying platform overheads.

In prior IaaS cloud performance evaluation research, reverse-engineering cloud providers was a common theme and lead to exploitation approaches for hardware heterogeneity~\citep{farley:12,ou:12}.
However, as hardware heterogeneity became less relevant over time~\citep{scheuner:18-cloud}, we refrain from interpreting the current lack of infrastructure inspection studies as a research gap that requires more attention.
The lack of studies from grey literature might also hint that this characteristic is currently of less interest to practitioners.

\subsection{Used Platform Configurations (\ref{rq:config})}\label{sec:configurations}
To answer \ref{rq:config}, we present a facetted barplot (Figure~\ref{fig:configurations}), visualizing the share of studies using a given configuration.
We report the share as percentage against all academic and all grey literature studies.

\subsubsection{Used Language Runtimes (\ref{rq:config_runtimes})}

\paragraph{Description}
The language runtime is the execution environment of a FaaS function.
Fully managed platforms offer a list of specific runtimes (e.g., Node.js, Python) determining the operating system, programming language, and software libraries.
Some providers support the definition of custom runtimes by following a documented interface, often in the form of Docker images.
If customization is not available in a platform, shims can be used to invoke an embedded target language runtime through a support runtime via system calls (e.g., invoking binaries through Node.js).

\begin{figure}
    \centering
    \includegraphics[width=0.47\textwidth]{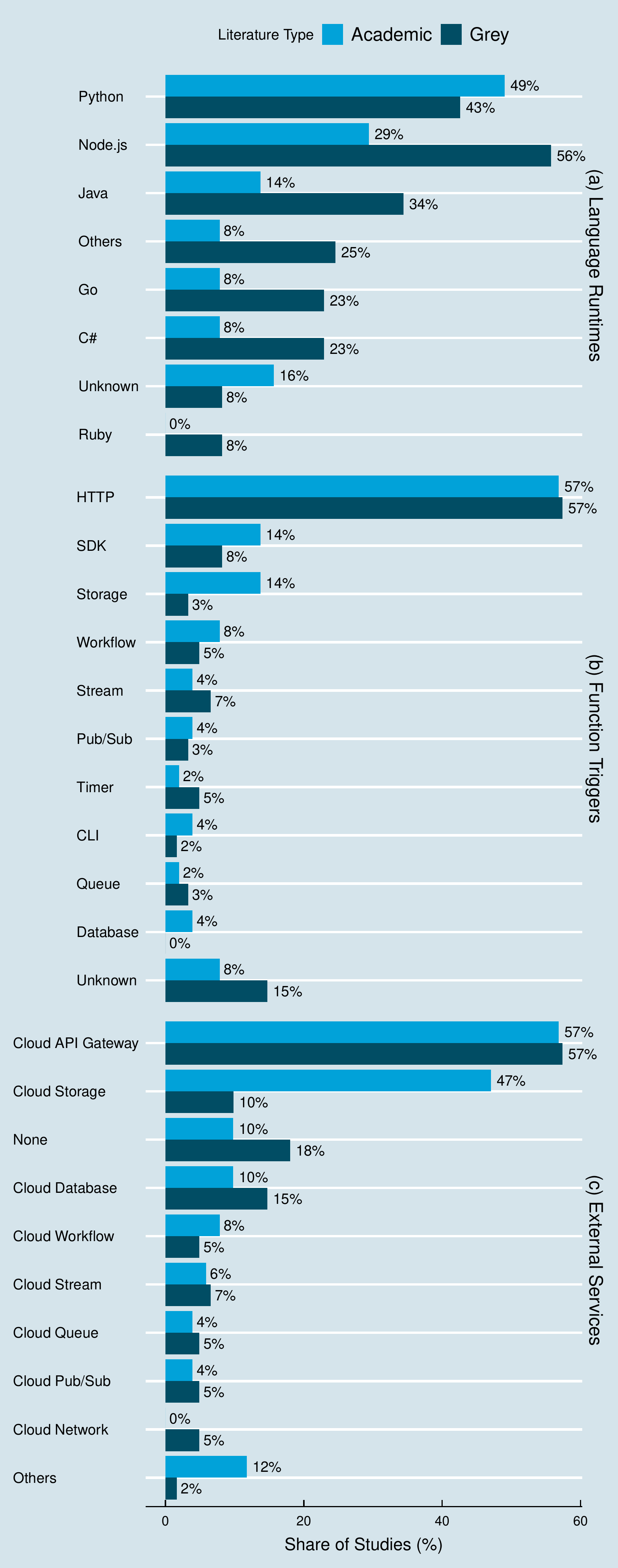}
    \caption{Used platform configurations for 51 academic and 61 grey literature studies}\label{fig:configurations}
\end{figure}

\paragraph{Results}
Figure~\ref{fig:configurations}a shows how frequently different language runtimes are evaluated.
Overall, Python and Node.js are the most popular runtimes, followed by Java. %
Interestingly, Node.js and Java are twice as popular among grey literature compared to academic literature. %
Grey literature generally covers more (up to 7), and more diverse languages in comparison to academic literature.
In particular, 46\% of the grey literature studies use more than one language in comparison to only 16\% for academic studies.
The category of \emph{Others} includes a list of 13 languages (e.g., F\#, Scala, Haskell) evaluated through custom runtimes or shims.

\paragraph{Discussion}
The large differences between academic and grey literature indicate a potential mismatch of academic and industrial interests.
This assumption is supported by other studies reporting the use of FaaS languages that similarly conclude Node.js to be roughly 20\% more popular than Python\footnote{Question 25 in \url{https://www.nuweba.com/blog/serverless-community-survey-2020-results}}~\citep{leitner:19}.

\subsubsection{Used Function Triggers (\ref{rq:config_triggers})}

\paragraph{Description}
Function triggers cover alternative ways of invoking FaaS functions.
FaaS functions can be triggered explicitly (e.g., through code) or implicitly through events happening in other services (e.g., image uploaded to cloud storage triggers function).
HTTP triggers invoke functions on incoming HTTP requests.
SDK and CLI triggers use software libraries to explicitly invoke functions.
Storage and database triggers invoke functions on object modifications (e.g., creation of a new object).
Stream, publish/subscribe (pub/sub), and queues process messages and invoke functions according to certain order or batching strategies.
Timer triggers invoke functions on specified schedules (e.g., cron expressions).
Workflow engines implement some kind of state machine to orchestrate complex processes across multiple functions.

\paragraph{Results}
Figure~\ref{fig:configurations}b shows how frequently different types of function triggers are evaluated.
HTTP triggers are by far the most commonly evaluated type of trigger and are used by 57\% of all studies.
Invocation through storage triggers is surprisingly uncommon for grey literature (10\%).
In general, only two studies cover more than two trigger types~[\ref{a41}, \ref{a27}], with the vast majority focusing on a single type of trigger.

\paragraph{Discussion}
It appears function triggering has received little attention given that most studies go for the de-facto default option of exposing a function via HTTP.
There are a wide range of other ways to trigger function execution (e.g., through a message queue, data streaming service, scheduled timer, database event, an SDK, etc.), which are currently not widely used and evaluated. %
The overall strong focus on HTTP triggers makes it hard to derive any meaningful differences between academic and grey literature, except for a stronger academic focus on storage (+11\%) and database triggers (4\% vs 0\%).

\subsubsection{Used External Services (\ref{rq:config_services})}

\paragraph{Description}
We now discuss which external services are commonly used in FaaS performance evaluations.
Cloud API gateways offer a fully managed HTTP service, which is commonly used to trigger functions upon incoming HTTP requests.
Cloud storages offer object storage for blob data, such as images.
Cloud databases offer structured data storage and querying.
Cloud workflow engines manage the state of complex processes across multiple functions.
Cloud stream, cloud queue, and cloud pub/sub are different types of message processing services.
Cloud networks refer to configurable private network services, such as AWS Virtual Private Cloud (VPC).

\paragraph{Results}
Figure~\ref{fig:configurations}c shows how frequently different external services are used.
Cloud API gateways are the most commonly used external service, which is unsurprising given that most studies use HTTP events to trigger functions.
About half of the academic literature studies use cloud storage compared to only 10\% of grey literature studies.
Overall, database services are among the most popular integrations.
The \emph{Others} category includes caching services, self-hosted databases, and special services such as artificial intelligence APIs.
In general, given how central service ecosystems are to the value proposition of cloud functions, it is surprising how rarely FaaS benchmarking studies incorporate external services beyond API gateways.

\paragraph{Discussion}
The result from function triggers explains the strong emphasis on cloud API gateway services for both, academic and grey literature.
Most surprisingly, cloud storage receives very little attention in grey literature (\(-37\%\)) compared to academic literature.
This is in contrast to other studies, indicating that database services are more commonly used in conjunction with FaaS\footnote{Question 18 in \url{https://www.nuweba.com/blog/serverless-community-survey-2020-results}}~\citep{leitner:19}.
A possible explanation lies in the strong focus of grey literature on micro-benchmarks, which typically use no external services or only an API gateway for easy invocation.
We conclude that the integration of external services in FaaS performance evaluations in a meaningful way remains a gap in current literature.

\subsection{Reproducibility (\ref{rq:reproducibility})}\label{sec:reproducibility}

\paragraph{Description}
To evaluate the maturity of literature concerning reproducibility, we rely on recent work by \citet{papadopoulos:19}.
They propose eight methodological principles for reproducible performance evaluation in cloud computing, which we now summarize and apply to our corpus:
\begin{enumerate}
    \item[P1] \emph{Repeated Experiments:} Repeat the experiment with the same configuration and quantify the confidence in the final result.
    \item[P2] \emph{Workload and Configuration Coverage:} Conduct experiments with different (preferably randomized) workloads and configurations motivated by real-world scenarios.
    \item[P3] \emph{Experimental Setup Description:} For each experiment, describe the hardware and software setup, all relevant configuration and environmental parameters, and its objective.
    \item[P4] \emph{Open Access Artifact:} Publish technical artifacts related to the experiment including software (e.g., benchmark and analysis code) and datasets (e.g., raw and cleaned data).
    \item[P5] \emph{Probabilistic Result Description:} Describe and visualize the empirical distribution of the measured performance appropriately (e.g., using violin or CDF plots for complex distributions), including suitable aggregations (e.g., median, 95th percentile) and measures of dispersion (e.g., coefficient of variation also known as relative standard deviation).
    \item[P6] \emph{Statistical Evaluation:} Use appropriate statistical tests (e.g., Wilcoxon rank-sum) to evaluate the significance of the obtained results.
    \item[P7] \emph{Measurement Units:} For all the reported measurements also report the corresponding unit.
    \item[P8] \emph{Cost:} Report the calculated (i.e., according to cost model) and charged (i.e., based on accounted resource usage) costs for every experiment.
\end{enumerate}

\begin{figure}[htb]
    \centering
    \includegraphics[width=0.46\textwidth]{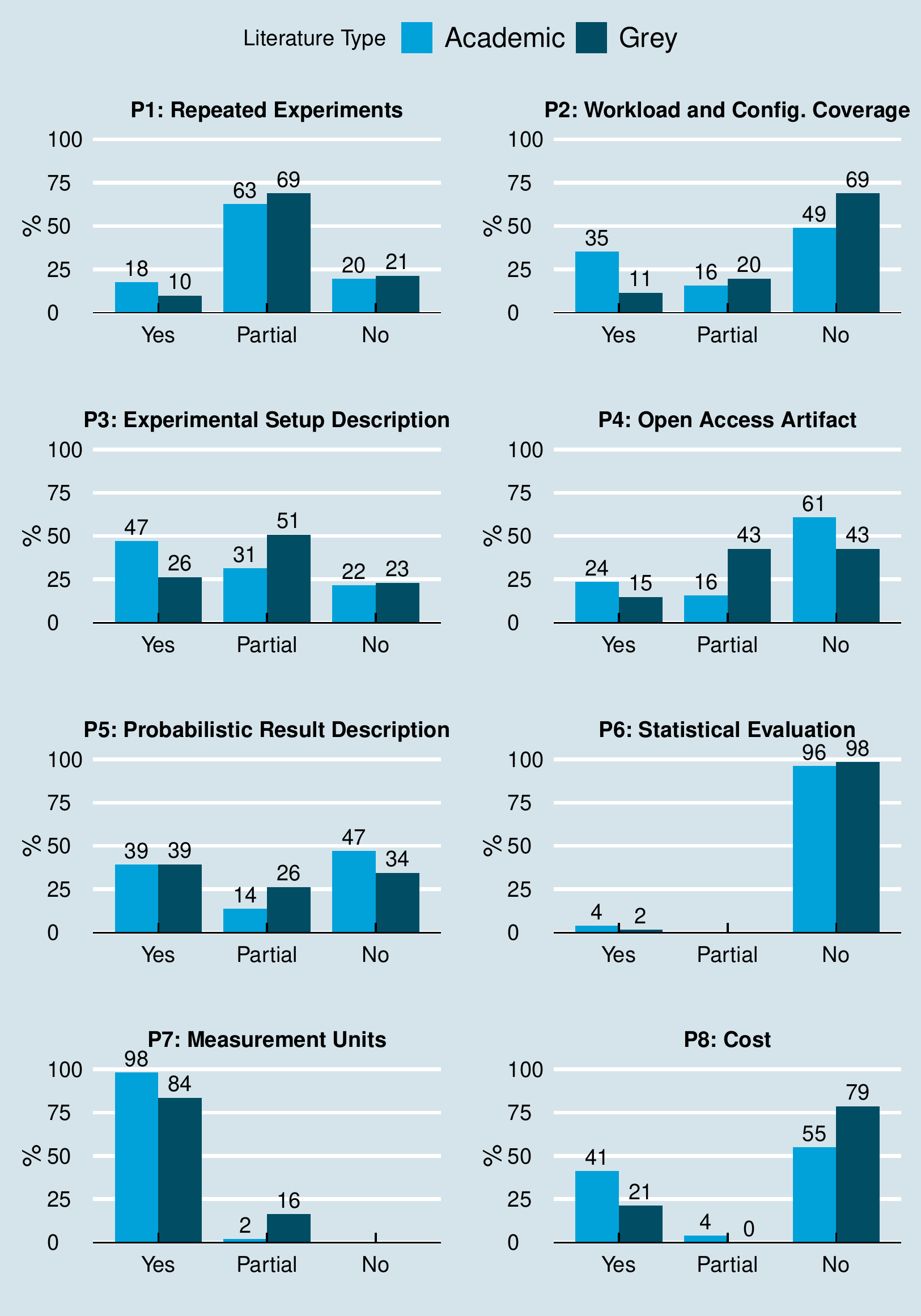}
    \caption{Evaluation of reproducibility principles P1--P8~\citep{papadopoulos:19} for 51 academic and 61 grey literature studies}\label{fig:reproducibility}
\end{figure}

\paragraph{Results}
Figure~\ref{fig:reproducibility} shows to what extent the reproducibility principles from \citet{papadopoulos:19} are followed by our selected academic and grey literature.
Overall, we find that 7 out of 8 principles are not followed by the majority of studies and, although academic literature performs better (\textgreater{}20\%) than grey literature for 3 principles (i.e., P2, P3, P8), we do not see a clear trend that academic work follows the proposed principles more strictly.
Interestingly, grey literature is even better than academic literature with regards to providing open access (P4) and probabilistic result descriptions (P5). %

\paragraph{P1: Repeated Experiments}\label{sec:p1_results}
The first subplot shows that the majority ($\approx$65\%) of our selected studies only partially fulfills P1 by performing some kind of repetition in their experiments, but without justifying or reporting confidence values.
The results for academic and grey literature are comparable.
While most of the studies perform some kind of repetition with the same configuration, the actual repetition count or duration time varies wildly. Results range from hundreds of repetitions or minutes of experimentation to thousands of repetitions or hours of experimentation collected over the course of multiple days.
Most of the studies partially fulfilling P1 do not discuss the confidence of their results. However, there are also a few individual cases that use appropriate probabilistic result descriptions (e.g., CDFs), but omit the absolute number of repetitions. %
Around 21\% of the studies perform insufficient repetitions (e.g., only tens of individual data points) or do not provide any information about whether or how their experiments use repetitions.
Only 10--15\% of the studies discuss the confidence of their results and collect a large-scale dataset with up to \num{360000} entries per histogram~[\ref{a08}] or up to \num{1000000} executions across 5 regions collected over 4 months~[\ref{g09}].
The following three studies further address the challenge of outdated benchmark results by offering a web service that continuously publishes updated results:
The $\lambda$ Serverless Benchmark\footurl{https://serverless-benchmark.com/} by Strehl~[\ref{g11}],
FaaSTest\footurl{https://www.faastest.com/} by the Nuweba company~[\ref{g57}],
and the cloud functions benchmarking dashboard\footurl{http://cloud-functions.icsr.agh.edu.pl/} by \citeauthor{figiela:18}~[\ref{a08}].

\paragraph{P2: Workload and Configuration Coverage}
About 50\% of the academic and 70\% of the grey literature studies do not use different workloads and configurations motivated by real-world scenarios.
These studies mostly use a single trivial example application, such as some version of a Hello World FaaS function\footurl{https://aws.amazon.com/getting-started/tutorials/run-serverless-code/} that returns a constant string, a timestamp, or a random value.
The second-most commonly used workload is some kind of CPU-intensive calculation, such as the prevalent prime number calculations with the iterative Sieve of Eratosthenes algorithm.
The partial category (16--20\%) mainly consists of studies using only micro-benchmarks or a single application-benchmark in only one configuration.
Application-level workloads motivated by real-world use cases and covering multiple configurations are used by 35\% of the academic but only 11\% of the grey literature studies.
One academic short paper by Kim and Lee~[\ref{a21}] specifically aims at introducing a suite of micro- and application-level serverless function benchmarks.

\paragraph{P3: Experimental Setup Description}\label{sec:p3_results}
More than half of all studies insufficiently describe their experimental setup.
However, as can be expected academic literature describes their experiments more comprehensively than grey literature.
About 31\% of the academic and 51\% of the grey literature studies omit important details in their experiment description and another 22\% of all studies exhibit severe flaws in their study description (e.g., significant details are missing), thus not fulfilling P3.
With only minor omissions (e.g., time of experiment not provided), 47\% of academic but only 26\% of grey literature studies satisfy P3.
The clear lead of academic literature highlights our overall impression that academic literature tends to describe their experimental setup in a more structured manner, often in dedicated sections (e.g., \citeauthor{kuhlenkamp:20}~[\ref{a26}]).
Further, academic studies tend to define their experiment goals more formally based on testable hypotheses (e.g., \citeauthor{figiela:18}~[\ref{a08}] or \citeauthor{manner:18}~[\ref{a34}]).

\paragraph{P4: Open Access Artifact}
Technical artifacts are unavailable for 61\% of the academic and 43\% of the grey literature studies.
Grey literature more commonly publishes their benchmark code (43\% vs 16\%) but more academic studies provide complete open source access to benchmark code and collected datasets (24\% vs 15\%).
The partial fulfillment category has only two exceptions of grey literature studies solely publishing their dataset but not their benchmark code instead of vice versa.

We discovered one of the following three practical issues related to handling open access artifacts in 9\% of all studies.
Firstly, we found inaccessible links in 3 studies that claim their artifacts are open source.
Secondly, we noticed obviously incomplete implementations (e.g., only for one provider, isolated legacy code snippet, code within inaccessible pictures) in another 3 studies.
Thirdly, we discovered open source artifacts that were not explicitly linked in 4 studies but could be discovered via manual Google or Github search or were implicitly linked in user comments (e.g., upon request of commenters).

The violation of P4 is particularly severe in combination with insufficient experimental setup description (P3).
A total of 19 (17\%) studies neither provide any technical artifacts nor any proper experimental setup description, rendering these studies practically impossible to replicate in practice.
Another 20 (18\%) studies violate P4 and omit relevant details in their experimental setup description.
Thus, these studies are hard to replicate under similar conditions (but a ``similar'' experiment could be conducted).

\paragraph{P5: Probabilistic Result Description}
About 40\% of all studies appropriately visualize or characterize their empirical performance data, but roughly the same percentage of all studies ignore complex distributions and primarily focus on reporting averages.
These nearly 40\% of the studies fulfilling P5 commonly use CDFs, histograms, or boxplots complemented with additional percentiles.
The 15\% of academic and 26\% of grey literature studies partially fulfilling P5 often give some selective characterization of the empirical distribution by plotting (raw) data over time or by violating P1 (i.e., insufficient repetitions).

\paragraph{P6: Statistical Evaluation}
Almost none of the selected studies perform any statistical evaluations.
Only two academic papers and one preprint use statistical tools such as the Spearman correlation coefficient~[\ref{a34}] or a nonparametric Mann-Whitney U test~[\ref{g49}].

\paragraph{P7: Measurement Units}
Overall, P7 is followed almost perfectly with no major violations.
Grey literature occasionally (16\%) omits measurement units (most commonly in figures) but the missing unit can be derived relatively easy from the context (or is mentioned in the text).

\paragraph{P8: Cost}
Cost models are missing in 55\% of the academic and 79\% of grey literature.
Two academic studies fulfill P8 partially by discussing costs in a general sense (e.g., as a motivational example), but without discussing actual costs of the experiments.
While some studies specifically focus on costs (e.g., \citeauthor{kuhlenkamp:17}~[\ref{a24}]), most studies typically calculate costs based on accounted or self-measured resource usage (e.g., runtime), but omit the actually charged cost.

\paragraph{Discussion}
We expected peer-reviewed academic literature to consistently achieve more methodological rigour than largely individually-authored grey literature.
Surprisingly, we do not see a clear trend that academic literature disregards the principles less often than grey literature. %
It is concerning that even simple principles such as publishing technical artifacts are frequently neglected, and grey literature is even better in providing at least partial open access.
Methodologically long-known principles from academic literature are still commonly overlooked in academia, exemplified by statistical guidance from 1986 on avoiding misleading arithmetic mean values~\citep{fleming:86}.
The presumably ``more informal'' grey literature is often on par or even better in appropriately describing performance results.

On the other hand, we emphasize that the clear lead of academic literature for three principles (i.e., P2, P3, P8) goes beyond the expressiveness of a 3-point discrete scale (i.e., yes, partial, no).
Experimental setup description (P3) has many facets and our results prevalently cover the presence or absence of relevant conditions, but fail to appropriately account for other important facets, such as clear structure and presentation.
Grey literature includes examples of unstructured studies, where results are presented without any discussion of methodology and scarce details about the experiment setup are scattered throughout a blog post.
In terms of P2, grey literature frequently picks one of the easiest available workloads, whereas academic studies more often motivate their workloads and attempt to link them to real-world applications.

We found that although many studies seemingly evaluate similar performance characteristics, comparing actual performance results is very difficult due to large parameter space, continuously changing environments, and insufficient experimental setup descriptions (P3).
We collected some exemplary results for the hosted AWS platform and find dramatic differences in numbers reported for platform overhead/cold starts ranging from 2ms (80th percentile, Python, 512mb but presumably reporting something else, maybe warm-start execution runtime of an empty function)~[\ref{g05}] up to 5s (median, Clojure via Java JVM, 256mb) [\ref{g54}].
More common results for end-to-end (i.e., including network latency of typically pre-warmed HTTPS connection) cold start overhead (i.e., excluding actual function runtime) for the Nodejs runtime on AWS (according to live data from 2020-02) are in the orders of $\approx$50ms (median) to $\approx$100ms (90th percentile)~[\ref{a08},\ref{g11}].
Studies from 2019 tend to report slightly higher numbers mostly around 200-300ms (median)~[\ref{g11},\ref{g33},\ref{g03}].

In the following, we highlight some insights into practical reproducibility related to P3 and P4.
We strongly agree with \citet{papadopoulos:19} that preserving and publishing experiment artifacts (P4) may be the only way to achieve practical reproducibility given that an exhaustive description (P3) of a complex experiment is often unrealistic.
We further argue that at least any time-consuming repetitive manual steps (but preferably any error-prone manual setup step that could lead to potential misconfiguration and affect the outcome of a study) should be fully automated~\citep{scheuner:14-cloudcom}.
We are positive to discover many automated setup and evaluation approaches in open source artifacts (P4) accompanying our studies, but still encounter too many studies with inexistent or tedious manual setup instructions.

\section{Implications and Gaps in Literature}\label{sec:gap}

We now discuss the main findings and implications of our study and identify gaps in current literature.

\subsection{Publication Trends (\ref{rq:trends})}

FaaS performance evaluation is a growing field of research in academic as well as grey literature, with a surge of new studies appearing in 2018.
Our results show that a healthy 20\% of the selected academic studies are published in top-ranked conferences or journals.

\subsection{Benchmarked Platforms (\ref{rq:platforms})}

The most evaluated platforms are AWS Lambda (88\%), Azure Functions (26\%), Google Cloud Functions (23\%), IBM Cloud Functions (13\%), and self-hosted platforms (14\%), predominantly Apache OpenWhisk. %
In absolute numbers, AWS is currently overstudied (by a factor of 3x).
However, other sources have reported that AWS is also predominant in actual production usage by a similar margin (see Section~\ref{sec:discussion_platforms}-Discussion).
Despite current industrial practice, future FaaS benchmarking studies should go beyond performance evaluations for the most popular platforms (e.g., avoid studying only AWS) to broaden our understanding of the field in general.
Further, other quickly rising cloud providers (e.g., Alibaba Cloud as the leading Asian cloud provider\footurl{https://www.parkmycloud.com/blog/alibaba-cloud-market-share/}) currently see no attention in literature.

\subsection{Evaluated Performance Characteristics (\ref{rq:characteristic})}

The lack of cross-platform benchmarks is a common theme across the following performance characteristics.

\subsubsection{Evaluated Benchmark Types (\ref{rq:characteristic_types})}
The predominant use of micro-benchmarks in 75\% of all studies indicates an over-emphasis on simple easy-to-build benchmarks, compared to application-benchmarks, which are used in 57\% of the academic and 31\% of the grey literature studies (i.e., overall 18\% use both).
This insight is supported by the large percentage of studies conducting platform overhead benchmarks with trivial functions (e.g., returning a constant) and CPU benchmarks using common workloads (e.g., prime number calculations).
Future work needs to go beyond such over-simplified benchmarks, and focus on more realistic benchmarks and workloads.
We also identify a need to develop cross-platform application-level benchmarks
as the current focus on a single platform (88\% of all application-benchmarks are evaluated on AWS) limits their usefulness for comparing platforms.
However, such cross-platform benchmarks are challenging to develop due to heterogeneous platforms and their complex ecosystems~\citep{eyk:20-hotcloudperf}.

\subsubsection{Evaluated Micro-Benchmarks (\ref{rq:characteristic_micro})}
Most micro-benchmarks (40\%) evaluate CPU performance, and show that CPU performance in FaaS systems is indeed proportional to the memory size of the selected function type for certain providers (i.e,. AWS, Google).
This is disappointing, as this behavior is well-documented by the cloud providers themselves and does not justify much further study.
We understand the need for periodic re-evaluations due to the dynamic nature of continuously evolving FaaS platforms~\citep{leitner:16} and want to emphasize the importance of studies targeting continuous benchmarking efforts (see examples in Section~\ref{sec:p1_results}-P1).
However, given the large scientific support that CPU performance of FaaS services behaves as documented, we suggest future studies to de-emphasize this aspect and focus on other characteristics such as network or function trigger performance (or real-world application-benchmarks).

\subsubsection{Evaluated General Characteristics (\ref{rq:characteristic_general})}

The most evaluated general performance characteristics are FaaS platform overhead (i.e., cold starts) and workload concurrency (i.e., invoking the same function in parallel), both used by about half of the studies.
This makes sense, as these aspects link to FaaS specifics and the most significant advantages of using FaaS, as reported in other surveys~\citep{leitner:19}.
No study currently evaluates function triggers across platforms.
We think the integration through triggers is an important aspect for FaaS performance, where insights can guide decisions about function invocation, function coordination, and usage of appropriate external services.
A major open research challenge towards such cross-platform benchmarks is the heterogeneous landscape of FaaS systems~\citep{eyk:20-hotcloudperf}. %

\subsection{Used Platform Configurations (\ref{rq:config})}

Our study indicates a broader coverage of language runtimes, but shows that other platform configurations focus on very few function triggers and external services.

\subsubsection{Used Language Runtimes (\ref{rq:config_runtimes})}

We identify a mismatch between academic and industrial sources, as Node.js, Java, Go, and C\# are evaluated two times more frequently in grey literature than in academic work.
Grey literature is generally more focused on covering more and more diverse runtimes than academic literature.
We suggest future academic literature studies to diversify their choice of runtimes, potentially also including insufficiently researched runtimes, such as Go or C\#.

\subsubsection{Used Function Triggers (\ref{rq:config_triggers})}

At the moment, a majority of studies (57\%) focuses on HTTP triggers.
We conclude that many trigger types remain largely insufficiently researched and suggest future studies to explore alternative triggers, such as message queues, data streams, timers, or SDKs.

\subsubsection{Used External Services (\ref{rq:config_services})}

Integrating external services in a meaningful way into FaaS performance evaluation studies remains an open challenge.
Despite their importance to overall serverless application performance, most current evaluations choose to abstract away from external services.
The only services we have seen used with some frequency are cloud API gateways (57\%), cloud storage (47\% academic vs 10\% grey literature), and cloud databases (10--15\%).

\subsection{Reproducibility (\ref{rq:reproducibility})}

We find that 7 of 8 reproducibility principles are not followed by the majority of the analyzed studies.
This is in line with the results of the original study~\citep{papadopoulos:19} on cloud experimentation in general.
We classify one third of all studies as practically impossible or hard to replicate under reasonably similar conditions due to the simultaneous lack of sufficient experimental setup description and available artifacts.
Overall, academic studies tend to satisfy the principles more comprehensively than grey literature but we do not see a clear trend that academic literature is less susceptible to disregarding the principles. %
Academic work is considerably better (principle fully met \textgreater{}20\%) than grey literature in choosing appropriate workloads (P2), describing the experimental setup (P3), and reporting costs (P8).
However, grey literature is considerably better in providing at least partial open access to experimental artifacts (i.e., code and data).
We support the trend towards artifact evaluations\footurl{https://www.acm.org/publications/policies/artifact-review-badging} and recommend focusing on artifact availability first (e.g., explicitly include availability check in reviewer guidelines) and subsequently target more qualitative attributes (e.g., ACM Functional, defined as documented, consistent, complete, exercisable). %
We conclude with actionable recommendations on what are the next steps regarding each principle for future FaaS studies to improve:
\begin{enumerate}
    \item[P1] Explicitly report the number of iterations. %
    \item[P2] Motivate workloads through industrial use cases.
    \item[P3] Report the time of experiment and follow good examples~[\ref{a26}, \ref{a08}, \ref{a34}] (see Section~\ref{sec:p3_results}–P3).
    \item[P4] Publish the dataset in addition to the benchmark code.
    \item[P5] Stop reporting mean values exclusively, but use appropriate statistical tools, such as CDFs, instead.
    \item[P6] Use appropriate statistical tests, such as Wilcoxon rank-sum or overlapping bootstrapped confidence intervals, for stronger conclusions~\citep{laaber:19-emse}.
    \item[P7] Include measurement units in all figures.
    \item[P8] Report a cost model. %
\end{enumerate}

\section{Related Work}\label{sec:relatedwork}
We compare and relate our results to existing literature reviews on FaaS and more generally on cloud performance evaluations, and compare our FaaS-specific results on reproducibility principles to the original study on cloud experimentation.

\subsection{Literature Reviews on FaaS}

\citet{kuhlenkamp:18} introduce a methodology for a collaborative SLR on FaaS benchmarking and report on preliminary results of 9 studies.
They capture more fine-grained experiments within each paper and extract data regarding workload generator, function implementation, platform configuration, and whether external services are used.
A completeness score of these categories represents the reproducibility of FaaS experiments.
Their results indicate insufficient experimental description.
\citet{somu:20} summarize the capabilities of 7 FaaS benchmarking studies along 34 characteristics for parameters, benchmarks, and metrics.
Their results indicate a strong focus on the AWS Lambda platform and a lack of support for function chaining, especially in combination with different trigger types.
These two most related papers hint towards some of our results but cannot confidently identify overall trends due to their limited scope.

\citet{taibi:20} conduct an MLR on serverless cloud computing patterns to catalogue 32 patterns originating from 24 sources.
Their MLR has a strong practitioner perspective but is limited to 7 peer-reviewed sources.
Our work focuses on performance whereas their pattern catalogue only occasionally mentions performance as part of discussing a pattern.

\citet{yussupov:19} conduct a systematic mapping study on FaaS platforms and tools to identify overall research trends and underlying main challenges and drivers in this field across 62 selected publications.
Their work covers a broader range of FaaS research and explicitly excludes FaaS benchmarking studies \enquote{without proposing any modifications}~\citep{yussupov:19} through their exclusion criteria.
Nevertheless, they identify 5 benchmarking studies and 26 function execution studies on performance optimization.
\citet{al-ameen:18} introduced a curated \enquote{Serverless Literature Dataset} that initially covered 60 scientific publications and preprints related to FaaS and Serverless computing in general, but in its latest Version 0.4 (2019-10-23)~\citep{spillner:19a} the dataset has been extended to 188 articles.
The authors classify their work as no survey itself, but rather envision its potential as input for future surveys such as ours.
We demonstrate this potential in the manual search process for academic literature where the serverless literature dataset covers 34 out of 35 relevant studies.
These two general studies identify publication trends, common technologies, and categories of research but do not extract and synthesize more specific data on FaaS benchmarking aspects we cover in our work.
To the best of our know\-ledge, we present the first comprehensive and systematic literature review on FaaS performance evaluation covering academic as well as grey literature.

\subsection{Literature Reviews on Cloud Performance}

We relate our results to existing literature reviews on general cloud performance topics.
These studies apply similar methods to us but in the context of cloud performance evaluation in general.
\citet{li:13a} conducted an SLR on evaluating commercial cloud services for 82 relevant studies.
Their work is methodologically closely related to our MLR but targets a more general field of research than our FaaS benchmarking study.
Their SLR has a strong focus on publication trends and performance metrics building upon the authors' previous work on cataloguing~\citep{li:12} and classifying~\citep{li:12b} performance evaluation metrics.
In contrast, our work specializes on performance characteristics in the field of FaaS, extends the scope beyond academic research by including grey literature, and reports on the reproducibility of the analyzed studies.
\citet{leitner:16} used an SLR methodology and open coding for identifying hypotheses seeding their principled experimental validation study on performance predictability of public IaaS clouds.
They performed experimental validation on common patterns of results and conclusions but did not extract further data on benchmarking studies.
A recent preprint (March 2020)~\citep{bjorndal:20} conducts an SLR on benchmarks and metrics within software engineering in the context of migrating from monolithic to microservice architectures.
The most frequent metrics for their 33 selected articles are latency, CPU, throughput, and network indicating that their study partially uses similar characteristics but in a less structured way (e.g., network and throughput are orthogonal aspects).

\subsection{Reproducibility Principles}

We compare our FaaS-specific results to the results of the original study by \citet{papadopoulos:19} on more general experimentation in cloud environments.
Our MLR study specifically targets FaaS experiments for academic and grey literature resulting in a largely disjoint set of studies with only 2 of our studies matching their stricter venue and impact criteria (i.e., $\geq$15  citations).
Overall, our results for academic literature studies are reasonably similar ($\pm$10\%) except for P1 and P5.
For P1, we speculate that we might have been more lenient in classifying studies, especially when no long-time experiments were present.
For P5, we see an improvement and notice more widespread use of CDFs, histograms, and boxplots or dotplots with error margins and accompanying percentiles.
Smaller trends suggest that more of our selected studies tend to open source technical artifacts (P4) and report costs (P8), but perform slightly worse in workload and configuration coverage (P2).

\section{Conclusion}\label{sec:conclusion}

This paper presented results from the first systematic and comprehensive survey on FaaS performance evaluation studies.
We conducted a multivocal literature review (MLR) across 112 studies from academic (51) and grey (61) literature.
We identify gaps in literature and give actionable recommendations highlighting the next steps towards compliance with reproducibility principles on cloud experimentation.
Our main findings are that AWS Lambda is the most evaluated FaaS platform (88\%), that micro-benchmarks are the most common type of benchmark (75\%), and that application benchmarks are currently prevalently evaluated on a single platform.
We further find that the majority of studies do not follow reproducibility principles on cloud experimentation from prior work.
Academic studies tend to satisfy the principles more comprehensively than grey literature, but we do not see a clear trend that academic literature is less susceptible to disregarding the principles.
We recommend future studies to broaden their scope of platforms beyond AWS as a single platform and in particular contribute cross-platform application-level benchmarks.
FaaS performance evaluation studies need to address flaws threatening their reproducibility and should particularly focus on choosing relevant workloads, publishing collected datasets, and statistically evaluating their results. %
Our survey consolidates existing work and can guide future research directions.
It provides a useful instrument for the systematic discovery of related studies and thus helps future research to relate and discuss their results in a wider context. %

\section*{Declaration of Competing Interests}

The authors declare that they have no known competing financial interests or personal relationships that could have appeared to influence the work reported in this paper.

\section*{CRediT authorship contribution statement}

\textbf{Joel Scheuner:} Conceptualization, Methodology, Software, Validation, Investigation, Data curation, Writing - Original Draft, Writing - Review \& Editing, Visualization
\textbf{Philipp Leitner:} Conceptualization, Methodology, Investigation, Writing - Review \& Editing, Supervision, Funding acquisition

\section*{Acknowledgments}
We thank the SPEC RG Cloud Working group for insightful discussions on this topic and Jan-Philipp Steghöfer for his valuable inputs.
This work was partially supported by the Wallenberg AI, Autonomous Systems and Software Program (WASP) funded by the Knut and Alice Wallenberg Foundation, and by the Swedish Research Council VR under grant number 2018-04127.

\bibliographystyle{elsarticle-num-names}
\biboptions{sort&compress}
\bibliography{faas-mlr}

\newpage

\section*{Appendix}

The appendix provides a complete list of all 112 studies analyzed as part of the multivocal literature review including hyperlinks to 51 academic (see Table~\ref{tab:academic_list}) and 61 grey (Table~\ref{tab:grey_list}) literature sources.
The complete curated dataset is available online for interactive querying\footurl{\appendixurl} in its latest version and published as a versioned dataset on Zenodo~\citep{scheuner:20-jss-dataset}.

\begin{table*}[t]
    \centering
    \footnotesize
    \caption{\label{tab:academic_list} Complete List of Analyzed Academic Literature Studies.}
    \begin{tabular}{@{}lrp{2.4cm}p{12cm}l@{}}
        \toprule
        \textbf{ID}              & \textbf{Ref}                          & \textbf{Authors}                   & \textbf{Article Title and Link (preferrably DOI, last accessed 2020-07-09)}                                                                                                                                                                                 & \textbf{Published}    \\ \midrule
        \alabel{a01} A01 & \citep{akkus:18}              & \citeauthor{akkus:18}            & \href{https://www.usenix.org/conference/atc18/presentation/akkus}{SAND: Towards High-Performance Serverless Computing}                                                                 & 2018-07 \\
        \alabel{a02} A02 & \citep{albuquerque-jr:17}     & \citeauthor{albuquerque-jr:17}   & \href{https://www.thinkmind.org/download.php?articleid=icsea_2017_9_30_10096}{Function-as-a-Service X Platform-as-a-Service: Towards a Comparative Study on FaaS and PaaS}             & 2017-10 \\
        \alabel{a03} A03 & \citep{back:18}               & \citeauthor{back:18}             & \href{https://doi.org/10.1007/978-3-319-99819-0_11}{Using a Microbenchmark to Compare Function as a Service Solutions}                                                                 & 2018-08 \\
        \alabel{a04} A04 & \citep{balla:20}              & \citeauthor{balla:20}            & \href{https://doi.org/10.1007/978-3-030-38651-1_21}{Tuning Runtimes in Open Source FaaS}                                                                                               & 2020-01 \\
        \alabel{a05} A05 & \citep{bardsley:18}           & \citeauthor{bardsley:18}         & \href{https://doi.org/10.1109/SmartCloud.2018.00012}{Serverless Performance and Optimization Strategies}                                                                               & 2018-09 \\
        \alabel{a06} A06 & \citep{bortolini:20}          & \citeauthor{bortolini:20}        & \href{https://doi.org/10.1007/978-3-030-33509-0_16}{Investigating Performance and Cost in Function-as-a-Service Platforms}                                                             & 2019-10 \\
        \alabel{a07} A07 & \citep{carreira:18}           & \citeauthor{carreira:18}         & \href{http://learningsys.org/nips18/assets/papers/101CameraReadySubmissioncirrus_nips_final2.pdf}{A Case for Serverless Machine Learning}                                              & 2018-12 \\
        \alabel{a08} A08 & \citep{figiela:18}            & \citeauthor{figiela:18}          & \href{https://doi.org/10.1002/cpe.4792}{Performance evaluation of heterogeneous cloud functions}                                                                                       & 2018-08 \\
        \alabel{a09} A09 & \citep{fouladi:17}            & \citeauthor{fouladi:17}          & \href{https://www.usenix.org/conference/nsdi17/technical-sessions/presentation/fouladi}{Encoding, Fast and Slow: Low-Latency Video Processing Using Thousands of Tiny Threads.}        & 2017-03 \\
        \alabel{a10} A10 & \citep{fouladi:19}            & \citeauthor{fouladi:19}          & \href{https://www.usenix.org/conference/atc19/presentation/fouladi}{From Laptop to Lambda: Outsourcing Everyday Jobs to Thousands of Transient Functional Containers}                  & 2019-07 \\
        \alabel{a11} A11 & \citep{gan:19}                & \citeauthor{gan:19}              & \href{https://doi.org/0.1145/3297858.3304013}{An Open-Source Benchmark Suite for Microservices and Their Hardware-Software Implications for Cloud \& Edge Systems}                      & 2019-04 \\
        \alabel{a12} A12 & \citep{gimenez-alventosa:19}  & \citeauthor{gimenez-alventosa:19} & \href{https://doi.org/10.1016/j.future.2019.02.057}{A framework and a performance assessment for serverless MapReduce on AWS Lambda}                                   & 2019-08 \\
        \alabel{a13} A13 & \citep{gupta:18}              & \citeauthor{gupta:18}            & \href{https://doi.org/10.1109/BigData.2018.8622139}{OverSketch: Approximate Matrix Multiplication for the Cloud}                                                                       & 2018-12 \\
        \alabel{a14} A14 & \citep{hall:19}               & \citeauthor{hall:19}             & \href{https://doi.org/10.1145/3302505.3310084}{An Execution Model for Serverless Functions at the Edge}                                                                                & 2019-04 \\
        \alabel{a15} A15 & \citep{hendrickson:16}        & \citeauthor{hendrickson:16}      & \href{https://www.usenix.org/conference/hotcloud16/workshop-program/presentation/hendrickson}{Serverless Computation with OpenLambda}                                                  & 2016-05 \\
        \alabel{a16} A16 & \citep{ishakian:18}           & \citeauthor{ishakian:18}         & \href{https://doi.org/10.1109/IC2E.2018.00052}{Serving Deep Learning Models in a Serverless Platform}                                                                                  & 2018-04 \\
        \alabel{a17} A17 & \citep{ivan:19}               & \citeauthor{ivan:19}             & \href{https://doi.org/10.3390/computers8020050}{Serverless Computing: An Investigation of Deployment Environments for Web APIs}                                                        & 2019-06 \\
        \alabel{a18} A18 & \citep{jackson:18}            & \citeauthor{jackson:18}          & \href{https://doi.org/10.1109/UCC-Companion.2018.00050}{An Investigation of the Impact of Language Runtime on the Performance and Cost of Serverless Functions}                        & 2018-12 \\
        \alabel{a19} A19 & \citep{jiang:17}              & \citeauthor{jiang:17}            & \href{https://doi.org/10.1007/978-3-319-69035-3_51}{Serverless Execution of Scientific Workflows}                                                                                      & 2017-10 \\
        \alabel{a20} A20 & \citep{jonas:17}              & \citeauthor{jonas:17}            & \href{https://doi.org/10.1145/3127479.3128601}{Occupy the cloud: distributed computing for the 99\%}                                                                                    & 2017-09 \\
        \alabel{a21} A21 & \citep{kim:19}                & \citeauthor{kim:19}              & \href{https://doi.org/10.1109/CLOUD.2019.00091}{FunctionBench: A Suite of Workloads for Serverless Cloud Function Service}                                                             & 2019-07 \\
        \alabel{a22} A22 & \citep{kim:19a}               & \citeauthor{kim:19a}             & \href{https://doi.org/10.1109/FAS-W.2019.00051}{Network Resource Isolation in Serverless Cloud Function Service}                                                                       & 2019-06 \\
        \alabel{a23} A23 & \citep{klimovic:18}           & \citeauthor{klimovic:18}         & \href{https://www.usenix.org/conference/atc18/presentation/klimovic-serverless}{Understanding Ephemeral Storage for Serverless Analytics}                                              & 2018-07 \\
        \alabel{a24} A24 & \citep{kuhlenkamp:17}         & \citeauthor{kuhlenkamp:17}       & \href{https://doi.org/10.1007/978-3-319-69035-3_48}{Costradamus: A Cost-Tracing System for Cloud-Based Software Services}                                                              & 2017-10 \\
        \alabel{a25} A25 & \citep{kuhlenkamp:19}         & \citeauthor{kuhlenkamp:19}       & \href{https://doi.org/10.1145/3344341.3368796}{An Evaluation of FaaS Platforms as a Foundation for Serverless Big Data Processing}                                                     & 2019-12 \\
        \alabel{a26} A26 & \citep{kuhlenkamp:20}         & \citeauthor{kuhlenkamp:20}       & \href{https://doi.org/10.1145/3341105.3373948}{Benchmarking Elasticity of FaaS Platforms as a Foundation for Objective-driven Design of Serverless Applications}                   & 2020-03 \\
        \alabel{a27} A27 & \citep{lee:18}                & \citeauthor{lee:18}              & \href{https://doi.org/10.1109/CLOUD.2018.00062}{Evaluation of Production Serverless Computing Environments}                                                                            & 2018-07 \\
        \alabel{a28} A28 & \citep{li:19}                 & \citeauthor{li:19}               & \href{https://doi.org/10.1145/3366623.3368139}{Understanding Open Source Serverless Platforms: Design Considerations and Performance}                                                  & 2019-12 \\
        \alabel{a29} A29 & \citep{lloyd:18}              & \citeauthor{lloyd:18}            & \href{https://doi.org/10.1109/IC2E.2018.00039}{Serverless Computing: An Investigation of Factors Influencing Microservice Performance}                                                 & 2018-04 \\
        \alabel{a30} A30 & \citep{lloyd:18a}             & \citeauthor{lloyd:18a}           & \href{https://doi.org/10.1109/UCC-Companion.2018.00056}{Improving Application Migration to Serverless Computing Platforms: Latency Mitigation with Keep-Alive Workloads}               & 2018-12 \\
        \alabel{a31} A31 & \citep{lopez:18}              & \citeauthor{lopez:18}            & \href{https://doi.org/10.1109/UCC-Companion.2018.00049}{Comparison of FaaS Orchestration Systems}                                                                                      & 2018-12 \\
        \alabel{a32} A32 & \citep{malawski:17}           & \citeauthor{malawski:17}         & \href{https://doi.org/10.1016/j.future.2017.10.029}{Serverless execution of scientific workflows: Experiments with HyperFlow, {AWS} Lambda and Google Cloud Functions} & 2017-11 \\
        \alabel{a33} A33 & \citep{malla:19}              & \citeauthor{malla:19}            & \href{https://doi.org/10.1002/itl2.137}{HPC in the Cloud: Performance Comparison of Function as a Service (FaaS) vs Infrastructure as a Service (IaaS)}                                & 2019-11 \\
        \alabel{a34} A34 & \citep{manner:18}             & \citeauthor{manner:18}           & \href{https://doi.org/10.1109/UCC-Companion.2018.00054}{Cold Start Influencing Factors in Function as a Service}                                                                       & 2018-12 \\
        \alabel{a35} A35 & \citep{mcgrath:16}            & \citeauthor{mcgrath:16}          & \href{https://doi.org/10.1109/CLOUD.2016.0060}{Cloud Event Programming Paradigms: Applications and Analysis}                                                                           & 2016-06 \\
        \alabel{a36} A36 & \citep{mcgrath:17}            & \citeauthor{mcgrath:17}          & \href{https://doi.org/10.1109/ICDCSW.2017.36}{Serverless Computing: Design, Implementation, and Performance}                                                                           & 2017-06 \\
        \alabel{a37} A37 & \citep{mohan:19}              & \citeauthor{mohan:19}            & \href{https://www.usenix.org/conference/hotcloud19/presentation/mohan}{Agile Cold Starts for Scalable Serverless}                                                                      & 2019-07 \\
        \alabel{a38} A38 & \citep{mohanty:18}            & \citeauthor{mohanty:18}          & \href{https://doi.org/10.1109/CloudCom2018.2018.00033}{An Evaluation of Open Source Serverless Computing Frameworks.}                                                                  & 2018-12 \\
        \alabel{a39} A39 & \citep{niu:19}                & \citeauthor{niu:19}              & \href{https://doi.org/10.1145/3307339.3343465}{Leveraging Serverless Computing to Improve Performance for Sequence Comparison}                                                         & 2019-09 \\
        \alabel{a40} A40 & \citep{oakes:18}              & \citeauthor{oakes:18}            & \href{https://www.usenix.org/conference/atc18/presentation/oakes}{SOCK: Rapid Task Provisioning with Serverless-Optimized Containers}                                                  & 2018-07 \\
        \alabel{a41} A41 & \citep{pelle:19}              & \citeauthor{pelle:19}            & \href{https://doi.org/10.1109/CLOUD.2019.00054}{Towards Latency Sensitive Cloud Native Applications: A Performance Study on AWS}                                                       & 2019-07 \\
        \alabel{a42} A42 & \citep{perez:19}              & \citeauthor{perez:19}            & \href{https://doi.org/10.1145/3297280.3297292}{A Programming Model and Middleware for High Throughput Serverless Computing Applications}                                               & 2019-04 \\
        \alabel{a43} A43 & \citep{pu:19}                 & \citeauthor{pu:19}               & \href{https://www.usenix.org/conference/nsdi19/presentation/pu}{Shuffling, Fast and Slow: Scalable Analytics on Serverless Infrastructure}                                             & 2019-02 \\
        \alabel{a44} A44 & \citep{puripunpinyo:17}       & \citeauthor{puripunpinyo:17}     & \href{https://doi.org/10.1109/INFCOMW.2017.8116416}{Effect of optimizing Java deployment artifacts on AWS Lambda}                                                                      & 2017-05 \\
        \alabel{a45} A45 & \citep{saha:18}               & \citeauthor{saha:18}             & \href{https://doi.org/10.1109/CLOUD.2018.00113}{EMARS: Efficient Management and Allocation of Resources in Serverless}                                                                 & 2018-07 \\
        \alabel{a46} A46 & \citep{shillaker:18}          & \citeauthor{shillaker:18}        & \href{https://conferences.inf.ed.ac.uk/EuroDW2018/papers/eurodw18-Shillaker.pdf}{A provider-friendly serverless framework for latency-critical applications}                            & 2018-04 \\
        \alabel{a47} A47 & \citep{singhvi:17}            & \citeauthor{singhvi:17}          & \href{https://doi.org/10.1145/3152434.3152450}{Granular Computing and Network Intensive Applications: Friends or Foes?}                                                                & 2017-11 \\
        \alabel{a48} A48 & \citep{spillner:18}           & \citeauthor{spillner:18}         & \href{https://doi.org/10.1007/978-3-319-73353-1_11}{FaaSter, Better, Cheaper: The Prospect of Serverless Scientific Computing and HPC}                                                 & 2018-12 \\
        \alabel{a49} A49 & \citep{wang:18}               & \citeauthor{wang:18}             & \href{https://www.usenix.org/conference/atc18/presentation/wang-liang}{Peeking behind the curtains of serverless platforms}                                                            & 2018-07 \\
        \alabel{a50} A50 & \citep{werner:18}             & \citeauthor{werner:18}           & \href{https://doi.org/10.1109/BigData.2018.8622362}{Serverless Big Data Processing using Matrix Multiplication as Example}                                                             & 2018-12 \\
        \alabel{a51} A51 & \citep{zhang:19}              & \citeauthor{zhang:19}            & \href{https://doi.org/10.1145/3304112.3325608}{Video Processing with Serverless Computing: A Measurement Study}                                                                        & 2019-06 \\ \bottomrule
    \end{tabular}
\end{table*}

\begin{table*}[t]
    \centering
    \footnotesize
    \caption{\label{tab:grey_list} Complete List of Analyzed Grey Literature Studies.}
    \begin{tabular}{@{}lp{2.7cm}p{12cm}l@{}}
        \toprule
        \textbf{ID}              & \textbf{Authors}                           & \textbf{Article Title and Link (last accessed 2020-03-10)}                                                                                                                                                                         & \textbf{Published}  \\ \midrule
        \glabel{g01} G01 & Adam Matan                        & \href{https://www.slideshare.net/adamatan/how-to-make-it-faaster}{How to Make It Fa(a)ster?}     & 2018-02-07 \\
        \glabel{g02} G02 & Adir Shemesh                      & \href{https://www.nuweba.com/blog/is-azure-functions-3.0-ready-for-production}{Is Azure Functions 3.0 Ready For Production?}                                                                                     & 2019-12-17 \\
        \glabel{g03} G03 & Alessandro Morandi                & \href{https://www.simplybusiness.co.uk/about-us/tech/2019/03/aws-lambda-cold-start-vpc/}{Investigating the effect of VPC on AWS Lambda cold-start}                                                               & 2019-03-28 \\
        \glabel{g04} G04 & Alex DeBrie                       & \href{https://www.alexdebrie.com/posts/aws-api-performance-comparison/}{AWS API Performance Comparison: Serverless vs. Containers vs. API Gateway integration}                                                   & 2019-02-20 \\
        \glabel{g05} G05 & Algirdas Grumuldis                & \href{http://kth.diva-portal.org/smash/record.jsf?pid=diva2\%3A1298864&dswid=9085}{Evaluation of "Serverless'' Application Programming Model : How and when to start Serverles}                                   & 2019-04-25 \\
        \glabel{g06} G06 & Andre Luckow and Shantenu Jha     & \href{https://arxiv.org/abs/1909.06055}{Performance Characterization and Modeling of Serverless and HPC Streaming Applications}                                                                                  & 2019-09-13 \\
        \glabel{g07} G07 & Andrew Hoang                      & \href{http://scholarworks.csun.edu/handle/10211.3/193100}{Analysis of microservices and serverless architecture for mobile application enablement}                                                               & 2017-06-13 \\
        \glabel{g08} G08 & Andrew Smiryakhin et al.          & \href{https://varteq.com/java-vs-nodejs-on-aws-lambda-benchmark-survey/}{Java vs NodeJS on AWS Lambda: Benchmark Survey}                                                                                         & unknown    \\
        \glabel{g09} G09 & Andy Warzon                       & \href{https://www.trek10.com/blog/lambda-cron/}{How Reliable is Lambda Cron?}                                                                                                                                    & 2017-01-06 \\
        \glabel{g10} G10 & Bernd Strehl                      & \href{https://medium.com/elbstack/the-largest-benchmark-of-serverless-providers-ac19b55750f4}{The largest benchmark of Serverless providers.}                                                                    & 2018-09-07 \\
        \glabel{g11} G11 & Bernd Strehl                      & \href{https://serverless-benchmark.com/}{$\lambda$ Serverless Benchmark (Serverless Benchmark 2.0)}                                                                                                                      & live       \\
        \glabel{g12} G12 & Bruno Schionato et al.            & \href{https://www.freecodecamp.org/news/what-we-learned-by-serving-machine-learning-models-using-aws-lambda-c70b303404a1/}{What We Learned by Serving Machine Learning Models Using AWS Lambda}                  & 2018-10-30 \\
        \glabel{g13} G13 & Can Tepakidareekul                & \href{https://medium.com/@manus.can/serverless-platform-comparison-google-cloud-function-vs-aws-lambda-8e060bcc93b4}{Serverless Platform Comparison: Google Cloud Function vs. AWS Lambda}                       & 2018-11-26 \\
        \glabel{g14} G14 & Cloudflare, Inc.                  & \href{https://www.cloudflare.com/learning/serverless/serverless-performance/}{How Can Serverless Computing Improve Performance? | Lambda Performance}                                                            & unknown    \\
        \glabel{g15} G15 & Driss Amri                        & \href{https://drissamri.be/blog/2019/05/19/minimize-aws-lambda-java-cold-starts/}{How to minimize AWS Lambda Java cold starts}                                                                                   & 2019-05-19 \\
        \glabel{g16} G16 & Erica Windisch                    & \href{https://read.iopipe.com/understanding-aws-lambda-coldstarts-49350662ab9e}{Understanding AWS Lambda Coldstarts}                                                                                             & 2017-02-09 \\
        \glabel{g17} G17 & Erwin Van Eyk                     & \href{https://repository.tudelft.nl/islandora/object/uuid\%3Aaf1407a8-6141-446c-828b-f3a4f5bf5786}{The Design, Productization, and Evaluation of a Serverless Workflow-Management System}                         & 2019-06-21 \\
        \glabel{g18} G18 & Frederik Willaert                 & \href{https://www.linkedin.com/pulse/serverless-promise-aws-azure-function-performance-scaling-willaert}{The Serverless Promise: AWS and Azure Function Performance and Scaling}                                 & 2018-01-14 \\
        \glabel{g19} G19 & Ingo Müller et al.                & \href{https://arxiv.org/abs/1912.00937}{Lambada: Interactive Data Analytics on Cold Data using Serverless Cloud Infrastructure}                                                                                  & 2019-12-02 \\
        \glabel{g20} G20 & James Randall                     & \href{https://www.azurefromthetrenches.com/azure-functions-significant-improvements-in-http-trigger-scaling/}{Azure Functions – Significant Improvements in HTTP Trigger Scaling}                                & 2018-03-09 \\
        \glabel{g21} G21 & James Randall                     & \href{https://www.azurefromthetrenches.com/azure-functions-significant-improvements-in-http-trigger-scaling/}{Azure Functions – Significant Improvements in HTTP Trigger Scaling}                                & 2018-03-09 \\
        \glabel{g22} G22 & Jannik Kollmann                   & \href{https://www.inovex.de/blog/serverless-architecture-aws-lambda/}{Serverless Architecture with AWS Lambda}                                                                                                   & 2018-02-06 \\
        \glabel{g23} G23 & Jim Conning                       & \href{https://medium.com/@jconning/aws-lambda-faster-is-cheaper-6bf32f58d741}{AWS Lambda: Faster Is Cheaper}                                                                                                     & 2017-01-26 \\
        \glabel{g24} G24 & John Chapin                       & \href{https://blog.symphonia.io/the-occasional-chaos-of-aws-lambda-runtime-performance-880773620a7e}{The Occasional Chaos of AWS Lambda Runtime Performance}                                                     & 2017-02-24 \\
        \glabel{g25} G25 & Josef Spillner                    & \href{https://blog.zhaw.ch/icclab/faas-function-hosting-services-and-their-technical-characteristics/}{FaaS: Function hosting services and their technical characteristics}                                      & 2019-10-31 \\
        \glabel{g26} G26 & Jun Sung Park                     & \href{https://eng.lifion.com/aws-lambda-triggers-kinesis-vs-sqs-1a1c78a86600}{AWS Lambda Triggers: Kinesis vs SQS}                                                                                               & 2019-05-06 \\
        \glabel{g27} G27 & Kevin S Lin                       & \href{https://kevinslin.com/aws/lambda_cold_start_idle/#}{Benchmarking Lambda’s Idle Timeout Before A Cold Start}                                                                                                & 2019-02-05 \\
        \glabel{g28} G28 & Leonardo Zanivan                  & \href{https://medium.com/criciumadev/serverless-native-java-functions-using-graalvm-and-fn-project-c9b10a4a4859}{Serverless Native Java Functions using GraalVM and Fn Project}                                  & 2018-06-24 \\
        \glabel{g29} G29 & Luke Demi                         & \href{https://blog.coinbase.com/benchmarking-aws-lambda-ca3cfb3c25cd}{Exploring serverless technology by benchmarking AWS Lambda}                                                                                & 2019-03-28 \\
        \glabel{g30} G30 & Mark Fowler                       & \href{https://medium.com/@shouldroforion/battle-of-the-serverless-part-2-aws-lambda-cold-start-times-1d770ef3a7dc}{Battle of the Serverless — Part 2: AWS Lambda Cold Start Times}                               & 2019-10-28 \\
        \glabel{g31} G31 & Matthieu Napoli                   & \href{https://mnapoli.fr/serverless-php-performances/}{Serverless and PHP: Performances}                                                                                                                         & 2018-05-24 \\
        \glabel{g32} G32 & Mikhail Shilkov                   & \href{https://blog.binaris.com/serverless-at-scale/}{Serverless at Scale: Serving StackOverflow-like Traffic}                                                                                                    & 2019-01-20 \\
        \glabel{g33} G33 & Mikhail Shilkov                   & \href{https://mikhail.io/serverless/coldstarts/big3/}{Comparison of Cold Starts in Serverless Functions across AWS, Azure, and GCP}                                                                              & 2019-09-26 \\
        \glabel{g34} G34 & Mikhail Shilkov                   & \href{https://mikhail.io/2018/10/azure-functions-v2-released-how-performant-is-it/}{Azure Functions V2 Is Released, How Performant Is It?}                                                                       & 2018-10-10 \\
        \glabel{g35} G35 & Mikhail Shilkov                   & \href{https://blog.binaris.com/from-0-to-1000-instances/}{From 0 to 1000 Instances: How Serverless Providers Scale Queue Processing}                                                                             & 2018-11-19 \\
        \glabel{g36} G36 & Mustafa Akin                      & \href{https://engineering.opsgenie.com/how-does-proportional-cpu-allocation-work-with-aws-lambda-41cd44da3cac}{How does proportional CPU allocation work with AWS Lambda?}                                       & 2018-01-25 \\
        \glabel{g37} G37 & Mustafa Akin                      & \href{https://engineering.opsgenie.com/run-native-java-using-graalvm-in-aws-lambda-with-golang-ba86e27930bf}{sing GraalVM to run Native Java in AWS Lambda with Golang}                                          & 2018-11-06 \\
        \glabel{g38} G38 & Nathan Malishev                   & \href{https://www.freecodecamp.org/news/lambda-vpc-cold-starts-a-latency-killer-5408323278dd/}{How to manage Lambda VPC cold starts and deal with that killer latency}                                           & 2018-07-14 \\
        \glabel{g39} G39 & Nathan Malishev                   & \href{https://levelup.gitconnected.com/aws-lambda-cold-start-language-comparisons-2019-edition-\%EF\%B8\%8F-1946d32a0244}{AWS Lambda Cold Start Language Comparisons, 2019 edition}                              & 2019-09-04 \\
        \glabel{g40} G40 & Paul Batum                        & \href{https://azure.microsoft.com/en-us/blog/processing-100-000-events-per-second-on-azure-functions/?ref=msdn}{Processing 100,000 events per second on Azure Functions}                                         & 2017-09-19 \\
        \glabel{g41} G41 & Pranjal Gupta and Shreesha Addala & \href{https://github.com/g31pranjal/g31pranjal.github.io/blob/master/assets/serverless-report.pdf}{Experimental Evaluation of Serverless Functions}                                                              & 2019-09-13 \\
        \glabel{g42} G42 & Ran Ribenzaft                     & \href{https://medium.com/@ranrib/the-right-way-to-distribute-messages-effectively-in-serverless-applications-f427e4229e67}{The right way to distribute messages effectively in serverless applications}          & 2018-03-31 \\
        \glabel{g43} G43 & Ran Ribenzaft                     & \href{https://medium.com/@ranrib/how-to-make-lambda-faster-memory-performance-benchmark-be6ebc41f0fc}{How to make Lambda faster: memory performance benchmark}                                                   & 2018-02-15 \\
        \glabel{g44} G44 & Remy Chantenay                    & \href{https://blog.travelex.io/blazing-fast-microservice-with-go-and-lambda-d30d95290f28}{Blazing Fast Microservice with Go and Lambda}                                                                          & 2018-02-28 \\
        \glabel{g45} G45 & Robert Vojta                      & \href{https://www.zrzka.dev/aws-journey-api-gateway-lambda-vpc-performance/}{AWS journey – API Gateway \& Lambda \& VPC performance}                                                                               & 2016-10-30 \\
        \glabel{g46} G46 & Rodric Rabbah                     & \href{https://www.ibm.com/cloud/blog/see-performance-details-cloud-functions}{How to see performance details for cloud functions (FaaS)}                                                                         & 2018-02-28 \\
        \glabel{g47} G47 & Ryan Chard et al.                 & \href{https://arxiv.org/abs/1908.04907}{Serverless Supercomputing: High Performance Function as a Service for Science}                                                                                           & 2019-08-14 \\
        \glabel{g48} G48 & Sachin Shrestha                   & \href{https://www.theseus.fi/handle/10024/227117}{Comparing Programming Languages used in AWS Lambda for Serverless Architecture}                                                                                & 2019-06-19 \\
        \glabel{g49} G49 & Sebastián Quevedo et al.          & \href{https://easychair.org/publications/preprint_download/FK9c}{Evaluating Apache OpenWhisk-FaaS}                                                                                                               & 2020-02-05 \\
        \glabel{g50} G50 & Tai Nguyen Bui                    & \href{https://medium.com/the-theam-journey/benchmarking-aws-lambda-runtimes-in-2019-part-i-b1ee459a293d}{Benchmarking AWS Lambda runtimes in 2019 (Part I)}                                                      & 2019-07-04 \\
        \glabel{g51} G51 & Tai Nguyen Bui                    & \href{https://medium.com/the-theam-journey/benchmarking-aws-lambda-runtimes-in-2019-part-ii-50e796d3d11b}{Benchmarking AWS Lambda runtimes in 2019 (Part II)}                                                    & 2019-07-15 \\
        \glabel{g52} G52 & Tim Nolet                         & \href{https://hackernoon.com/aws-lambda-go-vs-node-js-performance-benchmark-1c8898341982}{AWS Lambda Go vs. Node.js performance benchmark: updated}                                                              & 2018-01-15 \\
        \glabel{g53} G53 & Vlad Holubiev                     & \href{https://serverless.zone/my-accidental-3-5x-speed-increase-of-aws-lambda-functions-6d95351197f3}{My Accidental 3–5x Speed Increase of AWS Lambda Functions}                                                 & 2016-12-12 \\
        \glabel{g54} G54 & Wojciech Gawroński                & \href{https://pattern-match.com/blog/2018/10/18/functional-programming-in-serverless-world/}{Functional Programming in Serverless World}                                                                         & 2018-10-18 \\
        \glabel{g55} G55 & Yan Cui                           & \href{https://theburningmonk.com/2017/06/aws-lambda-compare-coldstart-time-with-different-languages-memory-and-code-sizes/}{aws lambda – compare coldstart time with different languages, memory and code sizes} & 2017-06-13 \\
        \glabel{g56} G56 & Yan Cui                           & \href{https://read.acloud.guru/how-long-does-aws-lambda-keep-your-idle-functions-around-before-a-cold-start-bf715d3b810}{How long does AWS Lambda keep your idle functions around before a cold start?}          & 2017-07-04 \\
        \glabel{g57} G57 & Yan Cybulski                      & \href{https://www.nuweba.com/blog/Introducing-faastest-and-faasbenchmark}{FaaS Benchmarking Made Easy: Introducing FaaStest.com and Faasbenchmark}                                                               & 2019-10-02 \\
        \glabel{g58} G58 & Yun Zhi Lin                       & \href{https://read.acloud.guru/comparing-aws-lambda-performance-of-node-js-python-java-c-and-go-29c1163c2581}{Comparing AWS Lambda performance of Node.js, Python, Java, C\# and Go}                              & 2018-03-08 \\
        \glabel{g59} G59 & Zac Charles                       & \href{https://medium.com/@zaccharles/net-core-3-0-aws-lambda-benchmarks-and-recommendations-8fee4dc131b0}{NET Core 3.0 AWS Lambda Benchmarks and Recommendations}                                                & 2019-10-25 \\
        \glabel{g60} G60 & Zack Bloom                        & \href{https://blog.cloudflare.com/serverless-performance-with-cpu-bound-tasks/}{Comparing Serverless Performance for CPU Bound Tasks}                                                                            & 2018-07-09 \\
        \glabel{g61} G61 & Zack Bloom                        & \href{https://blog.cloudflare.com/serverless-performance-comparison-workers-lambda/}{Serverless Performance: Cloudflare Workers, Lambda and Lambda@Edge}                                                         & 2018-07-02 \\ \bottomrule
    \end{tabular}
\end{table*}

\end{document}